\documentclass[11pt,a4paper]{report}
\usepackage{setspace}

\usepackage{amssymb,graphicx,color}
\usepackage{amsfonts}
\usepackage{latexsym}
\usepackage{amsmath}
\usepackage[a4paper, margin = 3cm, bottom = 2.5cm]{geometry}
\usepackage{float}
\usepackage{hyperref}
\usepackage{multirow}
\usepackage[table,xcdraw]{xcolor}
\usepackage{enumerate}

\usepackage{indentfirst}
\title{{\vspace{-14em} \includegraphics[scale=0.4]{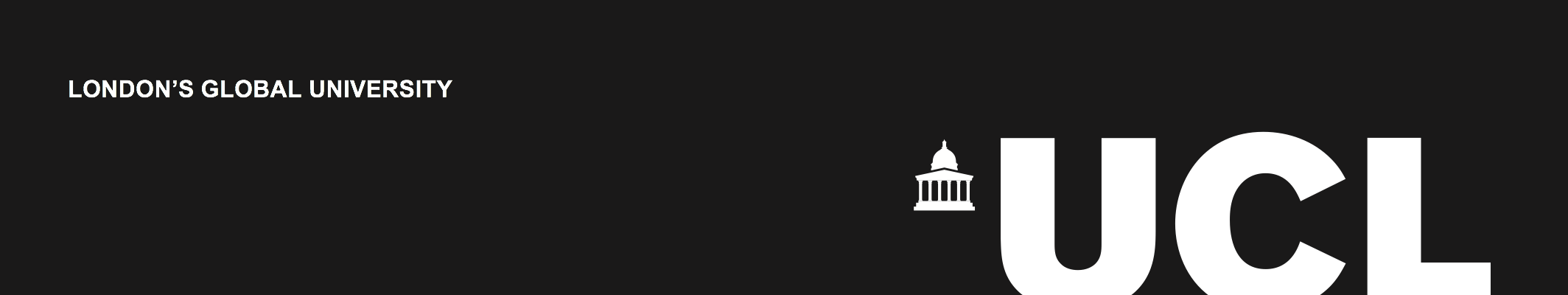}}\\
{{\Huge Genetic Analysis of Prostate Cancer with Computer Science Methods}}\\
}
\date{ }
\author{Yuxuan Li\thanks{
{\bf Disclaimer:}
This report is submitted as part requirement for the Computer Science Master's Degree at UCL. It is
substantially the result of my own work except where explicitly indicated in the text.
\emph{Either:} The report may be freely copied and distributed provided the source is explicitly acknowledged
\newline  
\emph{Or:}\newline
The report will be distributed to the internal and external examiners, but thereafter may not be copied or distributed except with permission from the author.}
\\ \\
Master of Engineering\\ \\
UCL Computer Science \\ \\
Supervised by Dr. Shi Zhou
}

\begin{document}

\maketitle
\begin{abstract}
Metastatic prostate cancer is one of the most common cancers in men. In the advanced stages of prostate cancer, tumours can metastasise to other tissues in the body, which is fatal. In this thesis, we performed a genetic analysis of prostate cancer tumours at different metastatic sites using data science, machine learning and topological network analysis methods. We presented a general procedure for pre-processing gene expression datasets and pre-filtering significant genes by analytical methods. We then used machine learning models for further key gene filtering and secondary site tumour classification. Finally, we performed gene co-expression network analysis and community detection on samples from different prostate cancer secondary site types. In this work, 13 of the 14,379 genes were selected as the most metastatic prostate cancer related genes, achieving approximately 92\% accuracy under cross-validation. In addition, we provide preliminary insights into the co-expression patterns of genes in gene co-expression networks. Project code is available at~\url{https://github.com/zcablii/Master_cancer_project}.
\end{abstract}
\tableofcontents
\setcounter{page}{1}

\chapter{Introduction}

\section{Problem Outline}
Metastatic prostate cancer, or prostate cancer, is a type of cancer that arises from the prostate gland and is caused by uncontrollable growth of some prostate cells. Globally, it is the most commonly diagnosed cancer with the exception of breast cancer and is the 5th most deadly cancer among men \cite{WHO14}. Prostate cancer is a malignant tumour in most cases and it can be difficult to control. Especially in the advanced stages of prostate cancer, the cancer cells can metastasise to other tissues in the body, such as the liver and lymph nodes, with the blood system and the lymphatic system. The cause of prostate cancer is largely unknown and there is no effective treatment to cure advanced prostate cancer. However, supportive treatment in the late stages can help patients control the progression of the disease and prolong their survival. \\

Prostate cancer typically develops very gently, and prostate cancer patients may be able to survive for years without symptoms or with mild symptoms. As the symptoms of prostate cancer are usually not obvious in the early stages, by the time a patient seeks treatment for severe symptoms, the patient is usually already at an advanced stage of prostate cancer. By this time the cancer has metastasised to other organs. For early stage prostate cancer, there is no effective way to know where the cancer will metastasise. If we can predict where the cancer will metastasise, we can use targeted treatments to prevent or mitigate cancer metastasis. This could significantly improve the survival rate of patients.\\

Recent medical research has shown that cancer is a genetic disease \cite{Vog04} and many cancer research institutes around the world have collected a large amount of cancer-related data, such as gene expression data of cancer cells from different cancer patients. However, most of these data have not been well analysed and explored. Therefore, researchers can take advantage of data science and computer science to perform genetic analysis on cancers, which can potentially explore the relationship between genes and cancers.\\

In complex diseases such as prostate cancer, the search for key disease genes has become an important task in scientific research in recent years. There are 2 main methods for filtering, studying and analysing cancer-related key genes: machine learning methods and complex network analysis methods. In recent years, there have been many studies showing that some machine learning models based on decision trees, such as random forests, work very well in dealing with problems in the field of bioinformatics \cite{Car06}, and therefore such models have become popular in the field of bioinformatics. The use of support vector machine (SVM) \cite{Cor95} models to predict cancer types from genetic alterations has also been found to be effective in much literature \cite{Soh17} \cite{Ram01}. Networked medical research using network science to abstract complex biological information into complex networks has also emerged as a typical disease research approach to understanding disease modules, identifying disease bio-markers and drug targets\cite{Ire16}.\\

\section{Contribution}
The main contributions of this project are threefold: firstly, we performed preliminary data cleaning and data analysis on a tumour cell gene expression dataset of prostate cancer patients. We proposed a general procedure for pre-processing gene expression datasets and pre-filtering of significant genes by purely analytical methods.\\

Secondly, we trained a high-performance decision tree-based machine learning model, XGBoost, to classify the site of metastasis in prostate cancer patients and achieved an accuracy of around 92\% with cross validation. We also discovered two sets of key genes which are related to prostate cancer metastasis, one with 13 genes and the other with 34 genes.\\

Finally, we performed gene co-expression network analysis and community detection in prostate cancer patients. The results indicate that there is a strong gene community structure of prostate cancer-related genes, and these gene communities may play different roles, leading to cancer metastasis to different secondary sites.\\

\section{Novel Methodology }
Although much of the literature has been devoted to genetic analysis of cancer using pure analytical methods, machine learning methods and gene co-expression network analysis methods respectively, there has been very little work combining them all together. The most outstanding methodology in our work is to take advantage of the strengths of these three approaches and combine them together.\\

In this project, in order to reveal the mystery between prostate cancer metastasis and genetics, both the “answer" to the question and the “reason" behind the “answer" are important. We hope to be able to accurately predict the likely site of cancer metastasis based on the patient's clinical data. But we also want to know which genes are factors in cancer and how they work together to influence the process of cancer metastasis. Because only by understanding the reasons can we have a better understanding in the life sciences. \\

The data analysis approach alone is not sufficient to achieve this goal, but it is useful for initially removing unrelated and unimportant information. The “black box" in machine learning models is suffering from interpretability issues \cite{Jap98}, and complex network models may not be as good at learning and prediction as machine learning models. In order to take advantage of each other's strengths and overcome each other's weaknesses, it is particularly important in this case that to ensemble machine learning and complex network models. \\

\chapter{Background}

\section{Prostate Cancer}
Cancer is one of the most deadly types of disease that mankind has yet to fully conquer. Cancer forms when cells are genetically altered in such a way that they are able to proliferate rapidly and indefinitely. This uncontrolled proliferation causes the tumour to grow larger and larger, and at a later stage it can even invade a secondary site for tumour development through the blood and lymphatic systems. This process of tumour spread is called metastasis. After a tumour cell has metastasised, the cancer cells from the original site will attach to the new body site and continue to grow. So essentially, the type of cancer cells in the new site is still the same as the cancer cells in the original site. And the tumour at the new site is known as a secondary tumour. It implies a secondary tumour is composed of prostate tumours, rather than LN tumours, if the prostate cancer metastasises to the lymph nodes (LN). The tumour in the LN is then referred to as a metastatic prostate cancer in the LN, rather than an LN cancer. In the following part of this thesis, we refer to prostate cancer patients whose tumours have metastasised to the LN, bone and liver as “LN patients”, “Bone patients” and “Liver patients” respectively. And we also refer to prostate cancer secondary site tumour samples which at LN, bone and liver as “LN samples”, “Bone samples” and “Liver samples” respectively.\\

Prostate cancer is a type of cancer that arises from the prostate gland. It is the 5th most deadly cancer among men\cite{WHO14}.In 2018, 1.2 million people were diagnosed with prostate cancer and nearly 359 thousand died from it \cite{Bra18} . It is a very common cancer in developed countries. Most prostate cancers grow slowly\cite{NCI14}, and may not cause any symptoms in the first few years. The greatest risk of prostate cancer is not the direct destruction of the prostate, but the possibility of the cancer rapidly metastasising to other body tissues and organs, particularly the bones and lymph nodes\cite{Rud15}. Metastatic prostate cancer can develop and grow rapidly, causing the organs in the secondary site to collapse quickly thus causing severe symptoms and hastening the patient's death.\\

\section{Gene Expression}
The genes in DNA encode information about protein molecules that are critical to the body's cells and perform functions essential to life. Different genes responsible for the production of different proteins, and different proteins perform different cellular functions in human body. In simple terms, to express a gene implies producing the corresponding protein for that gene. There are two main steps in the process of gene expression. First, the genetic messages in the DNA are transferred to the messenger RNA (mRNA) molecule by “transcription” \cite{Liu16}. At this point, the resulting mRNA has a copy of the genetic information in the DNA.  The cell then produces a protein molecule based on the genetic information on the mRNA by a process called “translation” \cite{Bio14}. So, if want to know what kind of protein a cell will make and how much, it can be found out by measuring the type and number of genes covered in the mRNA. And this measurement process is known as “RNA sequencing” \cite{RNA-Seq}. By sequencing, the kind of genes contained in the mRNA, and the``gene expression value" of each gene, can be obtained. Generally speaking, if a gene has a high gene expression value, then that gene is more expressed, which means that the protein corresponding to that gene will be produced in greater numbers.\\

Proteins expressed by a single gene usually do not have a specific function because of their simplicity. Often, multiple genes are involved in the synthesis of a functional and complex protein. The process of multiple genes working together to produce a complex protein is known as “gene co-expression”. \\

\subsection{Gene Co-expression Network (GCN)}
Typically, functional proteins are very complex and their synthesis often involves the co-expression of multiple genes. The gene co-expression network (GCN) \cite{Stu03} is a network used to represent gene co-expression information and relationships. In GCN, a point represents a gene, and if an obvious co-expression relationship exists between a pair of nodes, then a single undirected edge is used to connect them. The Pearson correlation coefficient can usually be used to measure the relationship between two gene co-expressions. Because co-expressed genes work together to synthesise a complex protein, the level of transcripts of co-expressed genes will increase and decrease together in different samples. GCNs can therefore be constructed from the correlation coefficients between genes. It is likely that there will be multiple obvious community structures in the GCN, where the genes in each community are likely to be co-expressed genes for certain proteins. \\

\chapter{Project overview}
\section{Project plan}
The rest of this thesis is organised as follows. In Chapter 4, we did a series of literature reviews in the field of cancer gene analysis. The literature includes pure analytical methods, machine learning methods and complex network methods for cancer-related gene selection and cancer prediction. Chapters 5 to 7 are the main body of this project, explaining the methods and the analysis of the results. Chapter 5 is concerned with dataset cleaning and pre-processing. In this section, we performed a thorough analysis of the original dataset and then proposed a general approach to pre-process the gene expression dataset. Chapter 6 focuses on key gene filtering and an ML model classification. In this section, we trained and validated on a XGBoost machine learning classification model and filtered out 2 key gene sets that were highly correlated with prostate cancer metastasis. In Chapter 7, we constructed gene co-expression networks for LN, Bone and liver patients, respectively, and then performed network analysis and community detection on these networks. We also compared these networks with each other to obtain further conclusions. Chapter 8 is a discussion of the results achieved, novelty contributions and limitations, and Chapter 9 is a guide to future work.\\

\section{Dataset}
The dataset used in this project is Metastatic Prostate Adenocarcinoma (SU2C/PCF Dream Team, PNAS 2019), which can be downloaded at cBioPortal \cite{cBioPortal}, a public website for cancer genomics. The dataset consists of the expression values of 19,293 genes from 270 patients. All gene expression values were determined by RNA sequencing of FPKM \cite{Soo11} standards.\\

Higher gene expression value of a gene means higher ``gene expression" of this gene in a tumour sample. Gene expression value of “0” means the measured value is zero. \\

The 270 patients covered LN, Bone, Liver, Prostate, Lung, Adrenal and Other Soft tissues patients. In detail, There are 117 LN patients, 74 Bone patients, 40 Liver patients, 7 Prostate patients, 6 Lung patients, 1 Adrenal patient and 25 Other Soft tissues patients.
\\

\chapter{Literature Review}
\section{Cancer Classification and Prediction }
\subsection{Analytical Methods }
In 1999, Golub T et al. \cite{Gol99} proposed an analytical method to differentiate acute myeloid leukaemia (AML) from acute lymphoblastic leukaemia (ALL) purely based on gene expression in bone marrow patient clinical samples. They analysed 6,817 human genes from bone marrow samples of 27 ALL and 11 AML patients and selected the top 50 most relevant genes based on the correlation between gene expression and disease type. Predictions were then made based on the differences in gene expression value between these genes in the two types of leukaemia. The results of this study proved in a mathematical way that the gene expression values of cancer-related genes differ between cancers and that it is feasible to classify cancers based on differences in the expression values of these genes. However, the study involved only 38 cases and their data covered just over 6,000 genes, which is limited compared to the total of over 20,000 genes in humans. They used only analytical methods, which may be effective when distinguishing only two cancers. But when it comes to multiple cancers classification, a purely mathematical analysis method alone may not be sufficient to solve the problem.\\

\subsection{Machine Learning Methods}
In 2001, Ramaswamy S et al. \cite{Ram01} proposed a machine learning approach based on gene expression values of tumour samples to classify a wide range of cancer cells. This study involved 218 tumour samples and 90 samples of normal human cells, covering up to 14 different cancers, and 16,063 genes. They used a support vector machine (SVM) algorithm to classify the samples and achieved an accuracy of 78\%. This study further demonstrates that it is possible to differentiate and diagnose cancers by differences in gene expression values of relevant genes between different cancers. They also found that the genes associated with different cancers tend to vary and are not fully captured by a few dozens of genes. Even though a few genes can be used to predict most cases, prediction can be more difficult for two cancers with similar tumour cells.\\

This study was further improved by Soh et al. \cite{Soh17} In this improved project, a larger dataset was used, which included 6640 samples involving 28 types of cancer. They used a support vector machine with recursive feature selection and eventually filtered out only 50 key genes and obtained an overall prediction accuracy of about 77.7\%.
\section{Gene Co-expression Network}

As mentioned in the backgroud section, information on gene co-expression is embedded in GCN, so a complex network analysis approach is also a way to explore cancer-related genes. In the work of Zhang, Bin and Horvath, Steve \cite{Zha05}, a general methodology for GCN analysis is proposed. This is a very important work as their proposed GCN has strong biological implication, which makes the results of the network analysis on this network valuable. This work focuses on the construction and feature analysis of GCNs. Details including the use of Pearson correlation as a co-expression measure for linkage weights, and then normalise edge weight values by applying different adjacency functions, including Hard Thresholding, sigmoid and power adjacency functions. They have also done preliminary work on gene co-expression community detection.\\

A recent work conducted by Savary, C et al. \cite{Sav20} in 2020 used the construction of GCNs and community detection method to identify cancer-related genes and to investigate which group of genes correspond to which type of cancer. They explored cancer-associated genes with the dataset which contains gene expression values from 820 paediatric cancer samples across six cancer group types. They constructed a weighted GCN where the linkage weight between each two genes was the absolute value of the Pearson correlation coefficient between these two genes in the samples. They analysed the community structure of this network and then to find the communities associated with each cancer by matching the correlation of the communities to the tumour type. These gene communities were then explored in depth to further filter out key genes associated with cancer.\\

\chapter{Data Pre-process and Analysis}

\section{Dataset Cleansing}
The dataset used in this project consists of the expression values of 19,293 genes from 270 patients. The 270 patients covered cases of metastatic prostate cancer to the LN, bone, liver, prostate, lung, adrenal and other soft tissues. However, for prostate, lung and adrenal patients, there are only a few cases. In order to avoid the specificity of a small number of data cases, this work focuses only on studying the categories with the majority of cases, i.e. LN, Bone and Liver patients. Therefore, the data under study includes 231 tumour samples from 117 LN patients, 74 bone patients and 40 liver patients, and each row is the gene expression value of one gene in all 231 tumour samples, and each column is the expression value of 19,293 genes in one tumour sample from one patient.\\

Then, we did follow steps to cleansing the dataset:

\begin{enumerate}[\hspace{3em}a.]
\item Remove any genes with gene expression values equals to 0 for all patients. 
\item   Remove any duplicate genes.
\item Reorder columns: Reorder the patients samples so that columns 1-117 are LN patients’ sample, 118-191 are Bone patients’ sample, and 192-231 are Liver patients’ sample.
\item All values are cut short to 3 digits after the decimal point.
\end{enumerate}
After the primary data cleaning process, 14,379 different genes remained, and this dataset was the main dataset for this project study. It is also referred to as the “original dataset” in the rest of this thesis.

\section{Dataset Analysis}
To better understand the characteristics of the gene expression values for all genes, we need to perform a thorough data analysis of the original dataset. The analysis consisted of examining the gene expression values for each gene, and the distribution characteristics of the maximum, minimum, mean and median values across the 231 patients.\\
\begin{figure}[h!]
\centering
\includegraphics[scale=0.5]{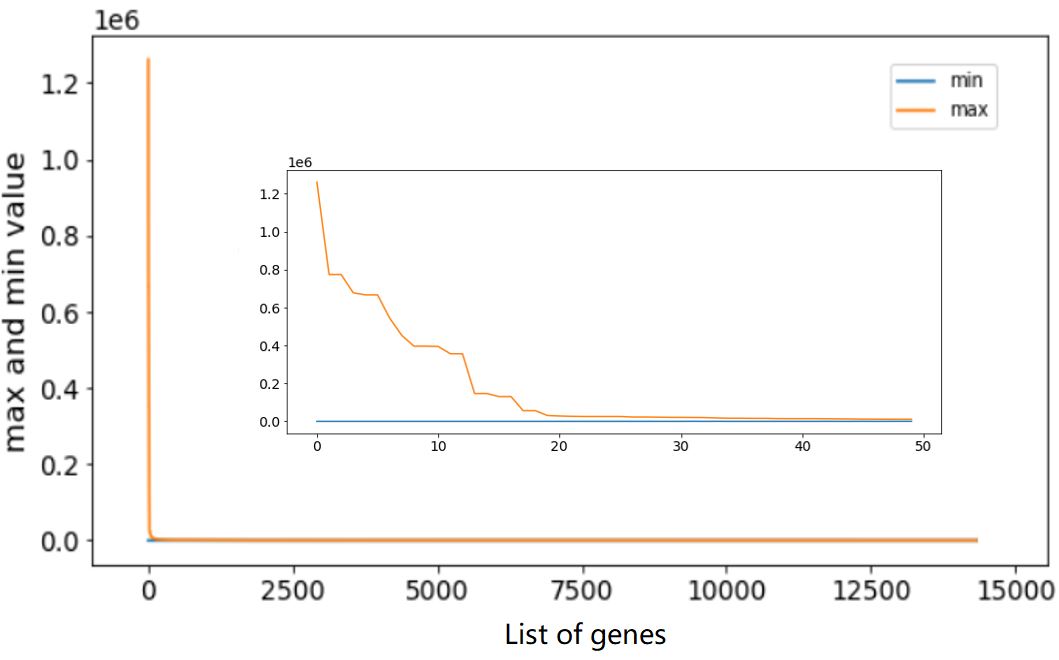}
\caption{\textit{The max and min gene expression values of the 231 patients for each of the 14,379 genes, where the genes are ordered in descending order of their Max value. The inner plot shows the top-50 ranked genes.}}
\label{fig:maxmin}
\end{figure}

\begin{figure}[h!]
\centering
\includegraphics[scale=0.5]{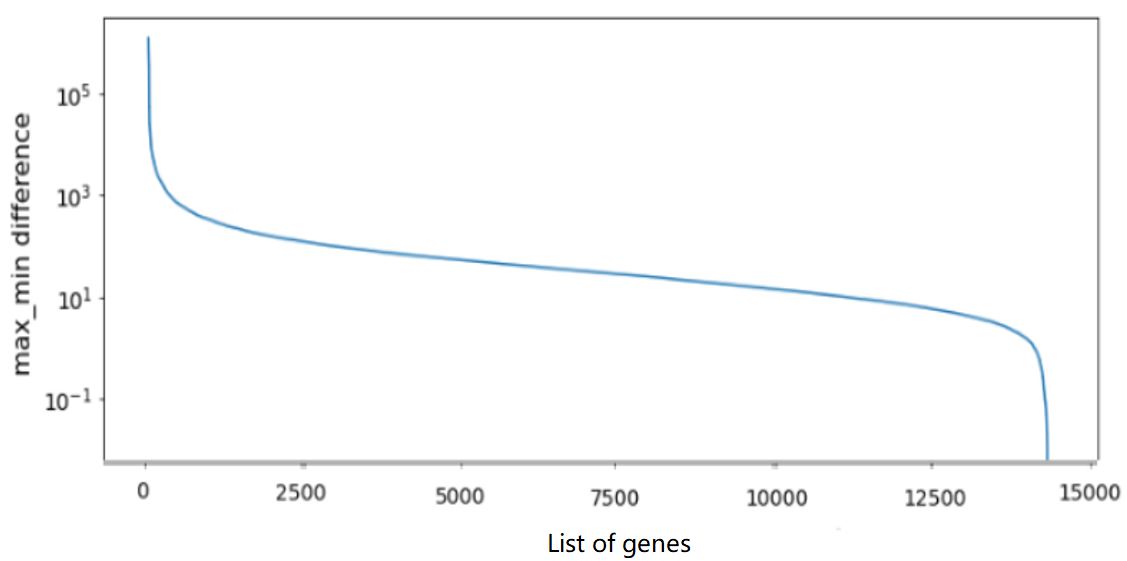}
\caption{\textit{The gene sensitivity of each gene, where genes are ordered in decreasing order of sensitivity value and sensitivity is in log scale. }}
\label{fig:sens}
\end{figure}
Figure \ref{fig:maxmin} shows the maximum and minimum expression values for each of the 14,379 genes in 231 patients. We can see that only about 20 genes had a very large difference between the maximum and minimum values. The range of gene expression values for most genes is relatively small. The difference between the maximum and minimum values of gene expression values is an informative indicator as it shows the sensitivity of a gene in all patients. In general, the higher the sensitivity of a gene, the easier it is for us to distinguish between patients with different cancer types. We then define the difference between the maximum and minimum gene expression values for a given gene as the ``gene sensitivity'' of that gene. As observed in Figure \ref{fig:sens}, most of the genes have gene sensitivity values in the range of 10 to 100. However, there are still some genes with extremely large or small gene sensitivities.Other than gene sensitivity, we are also investigated in the gene expression value distribution among the 14,379 genes. \\
\begin{figure}[h!]
\centering
\includegraphics[scale=0.9]{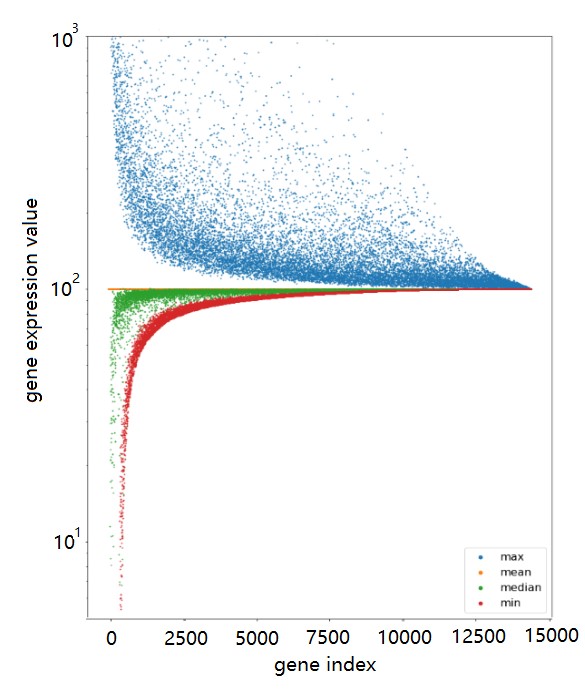}
\caption{\textit{Maximum, mean, median and minimum gene expression values for 14379 genes in 231 patients.  The genes are ordered in descending order of the mean value, Y-axis is in log scale. }}
\label{fig:geneval_distr}
\end{figure}

\begin{figure}[h!]
\centering
\includegraphics[width=0.8\textwidth]{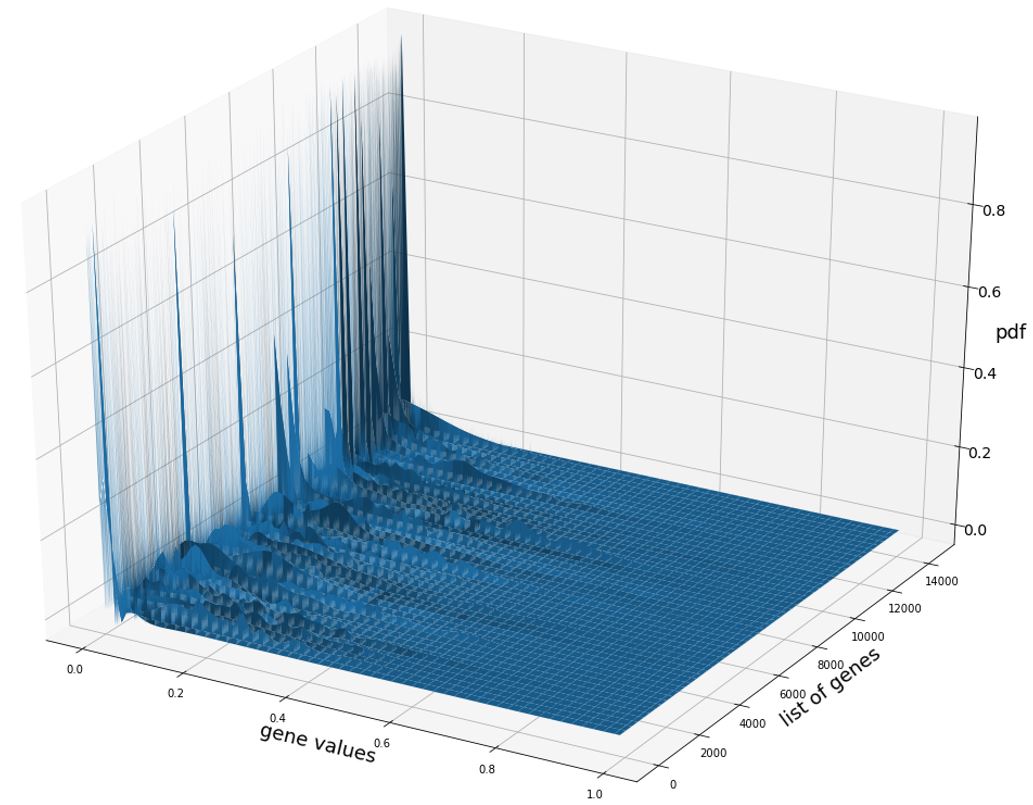}
\caption{\textit{Gene value probability distribution 3D plot: X: gene values linear mapping to range from 0 to 1 inclusive, and evenly divided into 100 bins; Y: all 14,379 genes; Z: the probability value of the gene Y. Genes are sorted in descending order of their sensitivity. }}
\label{fig:geneval_3d}
\end{figure}
In Figure \ref{fig:geneval_distr}, we note that the mean values for almost all genes are greater than the median, indicating a very uneven distribution of gene expression values, with most patients having very small gene expression values. For example, for one gene, a few patients had very large gene expression values, which drove up the mean value, but many patients had very small gene expression values so that the median gave a smaller value. To validate this thought, we added Figure \ref{fig:geneval_3d}, a 3D plot of the distribution of gene values for all genes. Clearly, this graph verifies that for most genes, many patients have very small gene expression values and only a few patients have very large gene expression values. \\
At this point, we have a general idea of the data set. For most genes, gene expression sensitivity varies widely, and gene expression values are very unevenly distributed.

\section{Pearson Correlation Coefficient}
One of the goals of this project is to identify cancer-related genes. Biological knowledge suggests that co-expressed genes work together to synthesise a complex protein and that the transcript levels of two co-expressed genes rise and fall together in different samples. Therefore, a direct study of the values of gene expression values is not very helpful for this work. In combination with biological knowledge, we believe that it would be more biologically meaningful to examine the trends and patterns in the rise and fall of gene expression values across samples. The Pearson correlation coefficient is an algorithm that describes the correlation between the fluctuating trends of two data. We therefore expected that the Pearson correlation coefficient for cancer-related gene expression values should be significantly higher for patients with the same type of prostate cancer secondary site than for patients with different types of prostate cancer secondary site. 

\begin{figure}[h!]
\centering
\includegraphics[scale=0.5]{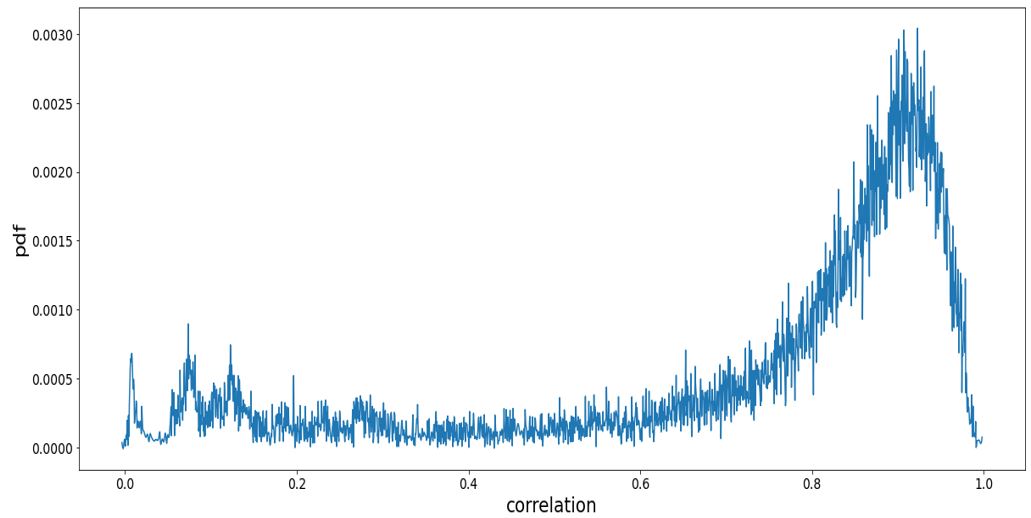}
\caption{\textit{Probability distribution function of Pearson correlation between all pairs of patients. } }
\label{fig:corrpdf}
\end{figure}

\begin{figure}[h!]
\centering
\includegraphics[scale=0.5]{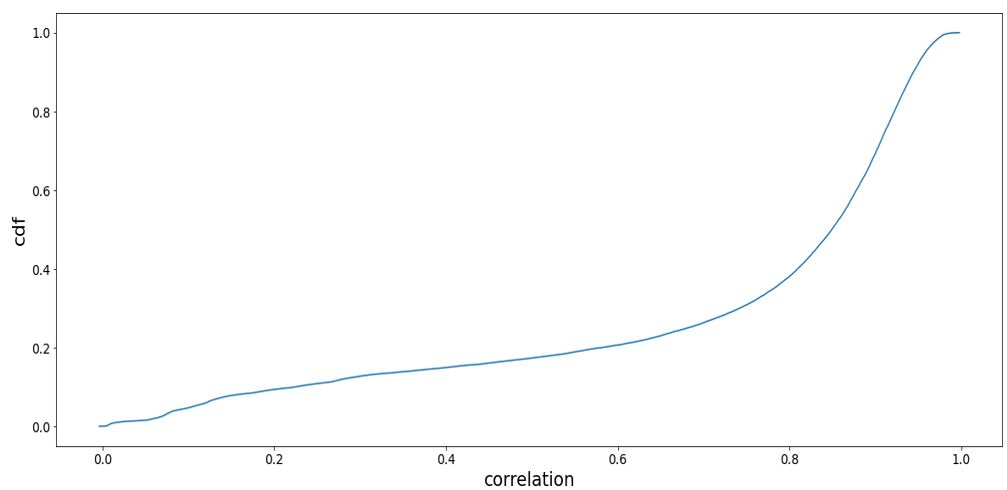}
\caption{\textit{Cumulative distribution function of Pearson correlation between all pairs of patients.}}
\label{fig:corrcdf}
\end{figure}

\begin{table}[]
\centering
\begin{tabular}{|l|l|l|l|llll}
\cline{1-4}
\begin{tabular}[c]{@{}l@{}}Mean of Pearson \\ Correlation\end{tabular} & LN & Bone & Liver &  &  &  &  \\ \cline{1-4}
LN & 0.818 & 0.68 & 0.802 &  &  &  &  \\ \cline{1-4}
Bone &  & 0.647 & 0.672 &  &  &  &  \\ \cline{1-4}
Liver &  &  & 0.825 &  &  &  &  \\ \cline{1-4}
\end{tabular}
\caption{\textit{Mean value of Pearson Correlation between patients of the same type of cancer (diagonal) and between patients of different types of cancer (off-diagonal). } }
\label{tab:meancorr}
\end{table}

\begin{figure}[h!]
\centering
\includegraphics[scale=0.6]{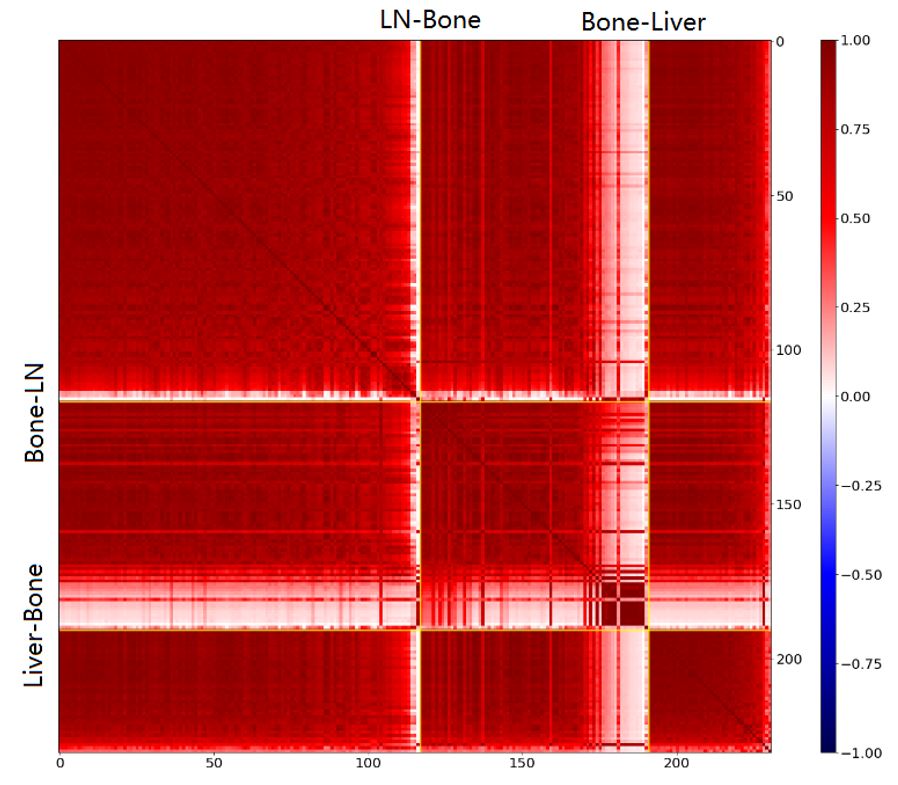}
\caption{\textit{Heatmap of Pearson coefficient of the 14,379 genes between the 231 patients. Patients of the same type of cancer are grouped together and sorted in descending order of their mean coefficient value within the group.}}
\label{fig:inihm}
\end{figure}

Figure \ref{fig:corrpdf} and \ref{fig:corrcdf} show the probability distribution function and cumulative distribution function of the Pearson correlation between all patient pairs. We found that most of the patients were highly correlated with each other, and the results in Table \ref{tab:meancorr} show that patients with the same type of cancer secondary site were highly correlated with each other, but patients with different types of cancer secondary site, such as LN and liver, were also highly correlated. This is also reflected in Figure \ref{fig:inihm}, where red indicates a high positive correlation, blue indicates a high negative correlation and white indicates no significant correlation. The square areas on the diagonal lines are correlations between patients with the same type of cancer, and the rectangles on the non-diagonal lines are correlations between patients with different types of cancer. We see that all the areas are in red, which indicates that most patients are highly correlated with each other, regardless of whether they have the same or different types of cancer secondary site patients. So it is not likely to effectively distinguish between different patients just based on all these 14,379 genes. This is probably because most genes that are not associated with cancer metastasis, which play an important role in the calculation of correlation values, are taken into account. In addition, since all 231 patient tumour cell samples were essentially prostate cancer cells with similar gene expression values, even though the cancer metastases related genes were expressed at different values in patients with different cancer metastasis types, they were so few in number compared to other common genes that the overall correlation would ignore these differences. \\

Therefore, the next step should be to filter out the irrelevant genes and keep only the genes associated with these three types of cancer metastases. With only genes associated with cancer metastasis, we expect only the diagonal squares in the heat map to be red and the other regions to be blue or white.

\section{Mask Selection Method}

To find genes that may be associated with cancer metastasis, we propose the mask selection method. In our understanding of cancer genetics, cancer metastasis related genes will have higher (cancer-causing genes) or lower (cancer-suppressing genes) gene expression values in patients with that type of cancer. Based on this idea, to filter out LN metastases related genes, we created an “LN metastases mask” for this dataset: a list of 231 elements, with the first 117 elements being 1 and the others being 0 (remembering that, in the dataset, the first 117 columns are gene expression values for LN patients). So, if a gene has a high positive correlation with the “LN metastases mask”, it is likely to be an LN site cancer-causing gene. And if a gene has a high negative correlation with the LN cancer mask, it is likely to be an LN cancer-suppressing gene. Similarly, we also created “Bone metastases mask” and “Liver metastases mask”. \\

LN metastases mask representation:

\[
  LN \ metastases \ mask = [\underbrace{1,..,1}_\text{117 1s}, \underbrace{0,..,0}_\text{114 0s} ]
\]

Bone metastases mask representation:  
 
\[
  Bone \ metastases \ mask = [\underbrace{0,..,0}_\text{117 0s}, \underbrace{1,..,1}_\text{74 1s} \underbrace{0,..,0}_\text{40 0s} ]
\]
Liver metastases mask representation: 

\[
  Liver \ metastases \ mask = [\underbrace{0,..,0}_\text{191 0s}, \underbrace{1,...,1}_\text{40 1s} ]
\]
\\

\begin{figure}[h!]
\centering
\includegraphics[scale=0.48]{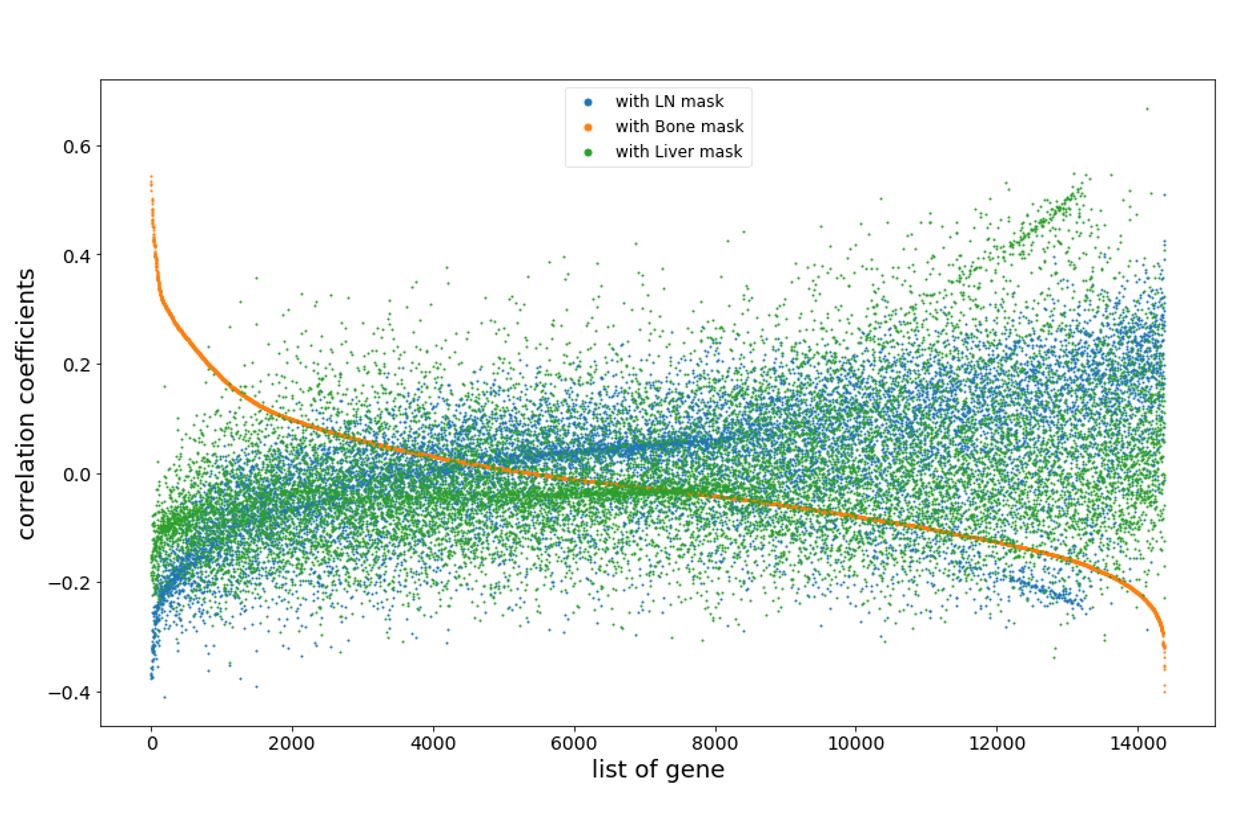}
\caption{\textit{X are 14,379  genes and Y are 3 correlation values with each of the LN, Bone and Liver metastases masks in different colours. Sort the genes in descending order according to the correlation value with the Bone metastases mask.}}
\label{fig:maskeff}
\end{figure}

To demonstrate the effectiveness of the mask selection method, we plotted Figure \ref{fig:maskeff}. The results in Figure \ref{fig:maskeff} show that the mask selection method is very effective, as we can easily select bone metastasis related genes based on this graph. We can clearly see that for genes with a high positive correlation with the bone metastasis mask, they have a low correlation or high negative correlation with the other two masks. And for genes with a high negative correlation with the bone metastasis mask, they have a low correlation or high positive correlation with the other two masks. These obvious gaps make them easy to distinguish.\\

The bone site cancer-causing genes are selected as follows: set a threshold value, e.g. 0.15, then any gene with a bone metastases mask correlation higher than 0.15 will be selected. Similarly, a threshold of -0.15 be set to select bone site cancer-suppressing genes in such a way that genes with the mask correlations below -0.15 would be selected. The same gene selection steps apply to LN and liver as well.\\

We have also sorted the above graph by descending order of correlation values with the LN and liver metastases masks respectively, which are put under Appendix section. All of these three plots give similar patterns. This suggests that we can use this masked selection method to pre-select genes that may be associated with prostate cancer metastasis.\\

\section{Normalisation}
As concluded in the ‘Dataset Analysis’ section, large variation in gene expression sensitivity and the uneven distribution of gene expression in the original dataset may introduce extra difficulties for data analysis and potentially affect the effectiveness of the mask selection method. Therefore, additional data processing of the dataset is necessary. In this section, we have attempted several normalisation schemes for gene expression values. The normalisation schemes we tried, along with an explanation of each scheme, are listed in Table \ref{Tab:norm}.\\

\begin{table}[]
\begin{tabular}{|l|l|llllll}
\cline{1-2}
\textbf{\begin{tabular}[c]{@{}l@{}}Normalisation \\ scheme\end{tabular}} & \textbf{Interpretation} &  &  &  &  &  &  \\ \cline{1-2}
Origin & Use the   original gene expression values &  &  &  &  &  &  \\ \cline{1-2}
Range & For   each gene, linear scale the original gene expression values to {[}0,1{]}. &  &  &  &  &  &  \\ \cline{1-2}
Log & \begin{tabular}[c]{@{}l@{}}For   each gene, take log\_10 value on all the 231 gene expression values, \\ and then   scale to range {[}0,1{]}.\end{tabular} &  &  &  &  &  &  \\ \cline{1-2}
Rank & \begin{tabular}[c]{@{}l@{}}Calculate   the gene sensitivity  values across the   231 patients. Then \\ replace values with ranks (between 1 and 231), same values   have the \\ same rank. Lastly scale the ranks into range {[}0,1{]}.\end{tabular} &  &  &  &  &  &  \\ \cline{1-2}
Logit & Take   logit value, i.e. y = -log(1/x -1) on the range-normalised values &  &  &  &  &  &  \\ \cline{1-2}
Logit-log & Take   logit value, i.e. y = -log(1/x -1) on the log-normalised values. &  &  &  &  &  &  \\ \cline{1-2}
\end{tabular}
\caption{\textit{Normalisation scheme interpretation.}}
\label{Tab:norm}
\end{table}

To verify and evaluate the performance of each normalization scheme, for gene values under each of the normalization schemes, do the following steps:

\begin{enumerate}[\hspace{3em}a.]
\item For each normalised gene (each gene has 231 data, one for each patient sample), calculate the correlation with each of the “LN metastases mask”, “Bone metastases mask” and “Liver metastases mask”.
\item Set a “threshold”, and if the absolute value of the correlation between a gene and any of the three masks is greater than or equal to the “threshold”, then the gene is retained. To find an optimal threshold, try thresholds in the range of 0.05 to 0.6 with a step size of 0.05.
\item Fill a table to record the number of genes selected by the three metastases masks with different thresholds. Then plot a graph to reflect this data. 
\end{enumerate}

Based on the results, we found that 0.15 and 0.2 were ideal thresholds because for each normalization scheme, these two thresholds selected a sufficient number of genes, and the positive and negative thresholds selected a relatively balanced number of genes. If a loose threshold like 0.1 is used, the number of genes considered will be too large, making this gene selection less effective. However, if a higher threshold is used, for example a threshold of 0.3, the number of genes selected will drop sharply to about only 1000. This may exclude some potentially informative gene candidates.\\

In the early stage, we want to take more genes into consideration, so we firstly use threshold of 0.15 instead of 0.20. To find out which normalisation scheme was most effective, we plotted a series of patient correlation heatmaps in which we set the threshold to 0.15 and calculated the correlation between any two patients for each normalisation scheme using only the union of the genes selected by LN, Bone and Liver cancer mask. \\

\begin{table}[]
\begin{tabular}{llllllll}
\cline{1-2}
\multicolumn{1}{|l|}{\includegraphics[scale=0.35]{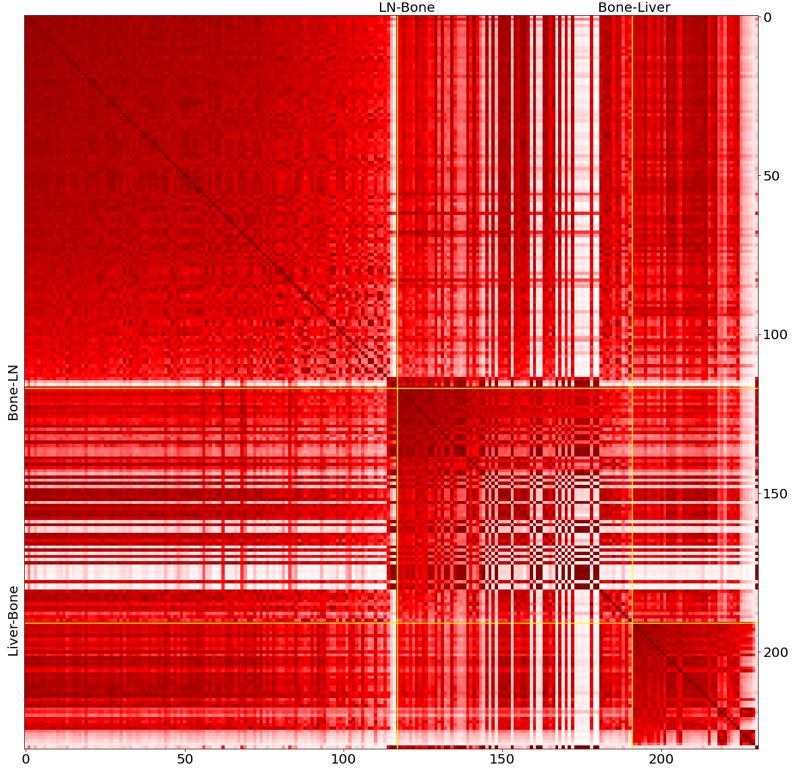}} & \multicolumn{1}{l|}{\includegraphics[scale=0.35]{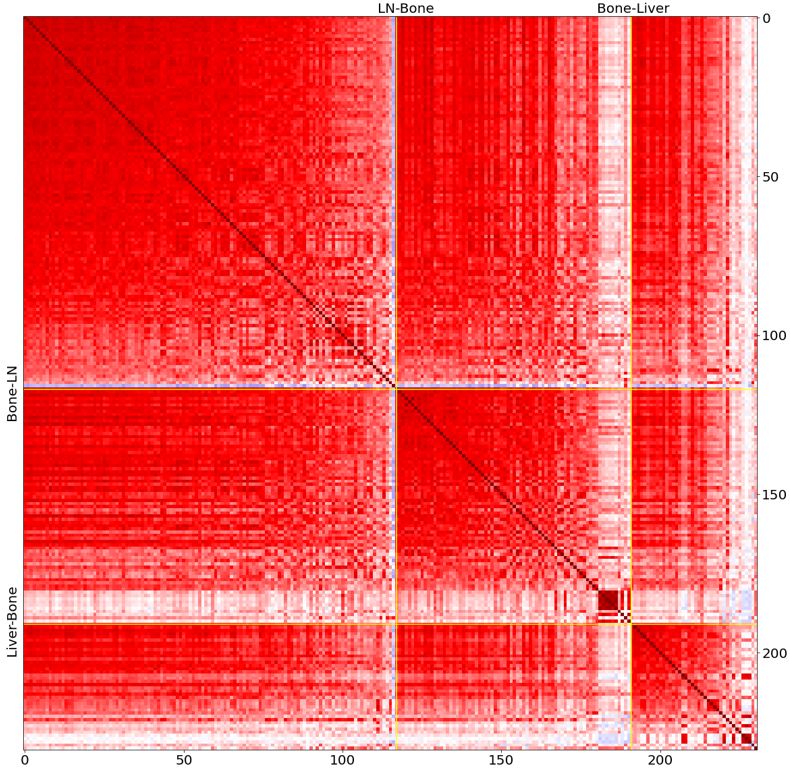}} &  &  &  &  &  &  \\ \cline{1-2}
\multicolumn{1}{|l|}{\includegraphics[scale=0.35]{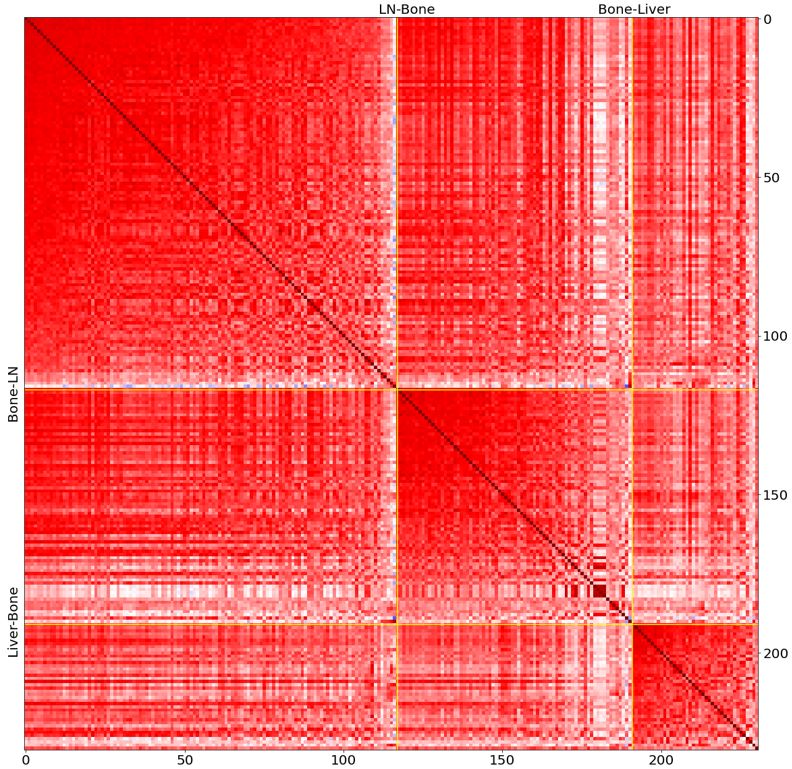}} & \multicolumn{1}{l|}{\includegraphics[scale=0.35]{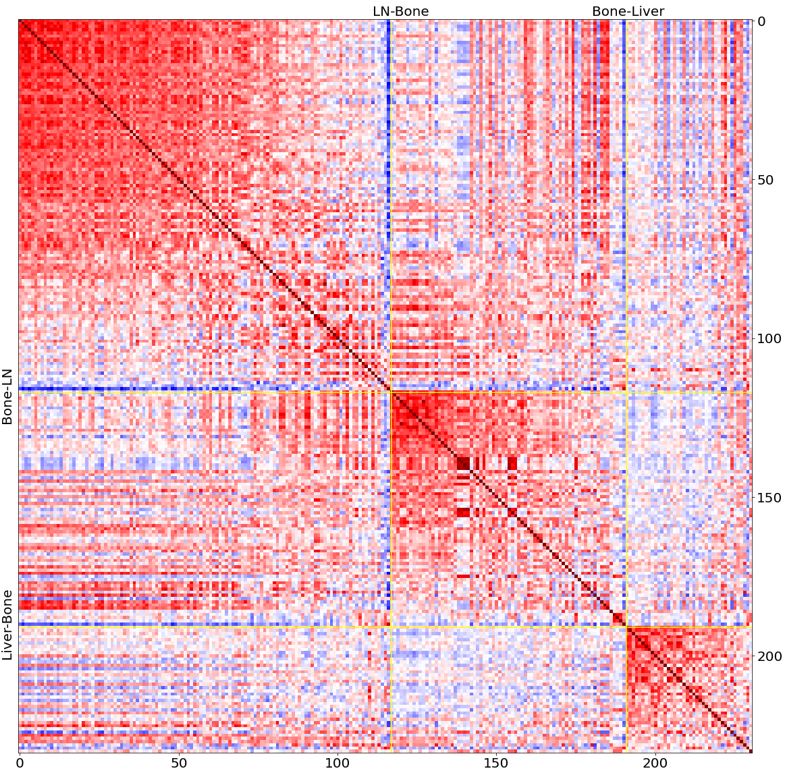}} &  &  &  &  &  &  \\ \cline{1-2}
\multicolumn{1}{|l|}{\includegraphics[scale=0.35]{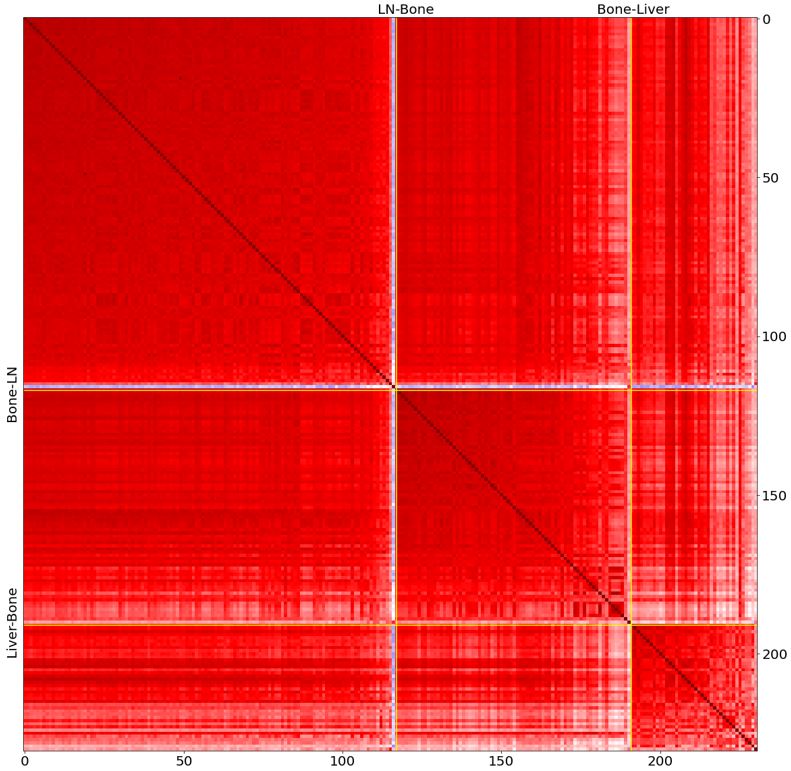}} & \multicolumn{1}{l|}{\includegraphics[scale=0.35]{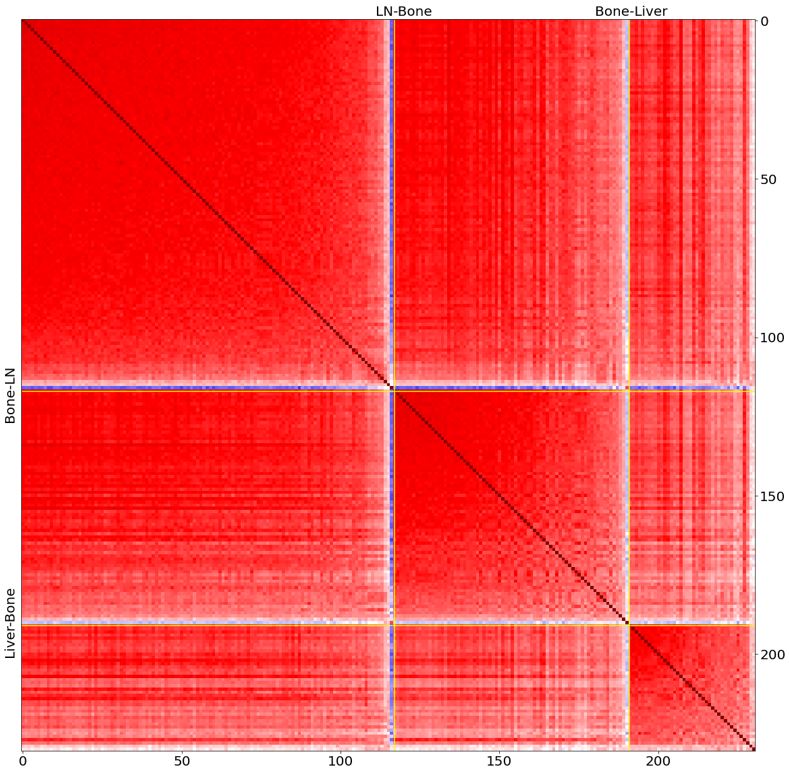}} &  &  &  &  &  &  \\ \cline{1-2}
\end{tabular}
\caption{\textit{Heatmaps of correlation between any of 231 patients’ selected genes where correlation is calculated using different normalised values: None, Range, log, rank, logit and logit-log normalisations, with threshold set to 0.15. Patients with the same kind of cancer are put in the same area and sorted in descending order of their mean coefficient value within the group.}}
\label{tab:hms}
\end{table}

Theoretically, if only cancer metastasis-related genes are involved, the ideal heat map would have only the diagonal squares in red and the other regions in blue or white. The results of the above experiments are added in Table \ref{tab:hms}. We can clearly see that in Table \ref{tab:hms}, the rank normalisation is the best normalisation scheme among the six schemes. Therefore, we will focus only on the rank normalisation of gene expression values.

\section{Gene Selection Strategies}
However, even with rank normalization and a threshold of 0.15, there is still too much red in the cross-section of the heatmap. This is probably because the threshold we used was too low, allowing many unrelated genes to be taken into account. Therefore, we should only consider genes that overlap because, as shown in Figure \ref{fig:maskeff}, if a gene is selected by a positive threshold of one cancer mask, it is likely to be selected by a negative threshold of another cancer mask as well. So the intersecting genes selected by the three masks are the ones that can be used to distinguish between different cancers. Based on this thought, we tried to use only the intersecting genes of the three mask selected, and refer these genes as “three masks intersect genes”.\\

\begin{figure}[h!]
\centering
\includegraphics[scale=0.6]{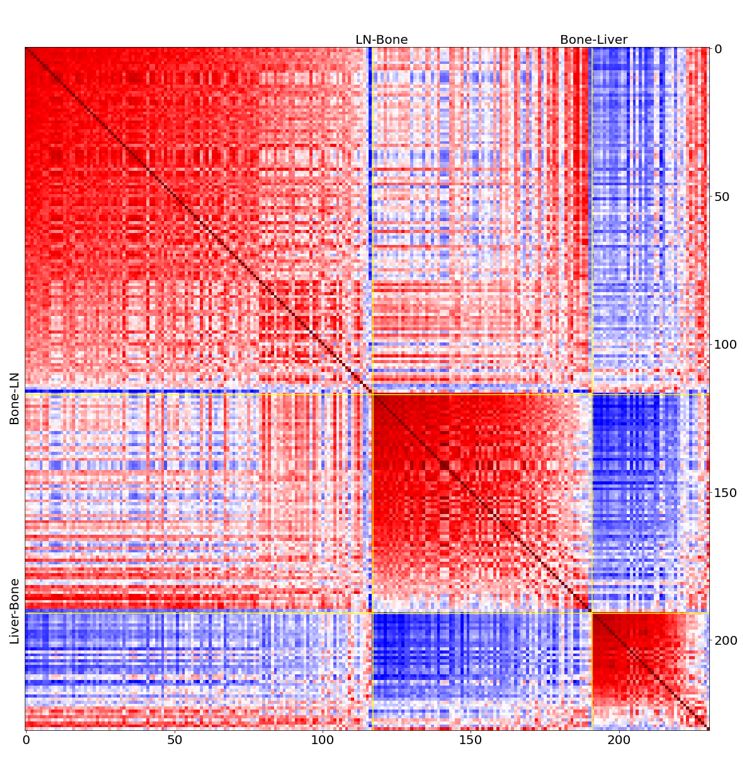}
\caption{\textit{Heatmaps of correlation between any of 231 patients’ rank normalisations genes selected by 3 all of metastases masks with threshold set to 0.15. There are 559 genes are taken into account.}}
\label{fig:559hm}
\end{figure}

The intersecting genes of the three masks selections with threshold of 0.15 gives 559 genes. These 559 genes are very informative and important genes, so we refer to these genes as SET\_559. This gene set will be used later in the chapter “Gene co-expression Network and Community Detection”. \\

We plotted the correlation heatmap of patients with the SET\_559 in Figure \ref{fig:559hm}, we note that liver metastases are easily detected from other site metastases, but the correlation between LN and bone patients remains high, making it difficult to distinguish between LN and bone patients based on the 559 genes currently selected.\\

Therefore, we conducted the following experiments to further search for a set of genes that could be used to differentiate between LN and bone patients. First, we ignored all liver patients and focused only on distinguishing LN and Bone patients. This time, instead of setting the threshold to 0.15, we chose a higher threshold of 0.2 to increase the differentiation of genes. We then selected genes with high positive correlation with LN mask and high negative correlation with Bone mask, as well as genes with high negative correlation with LN mask and high positive correlation with Bone mask. In other words, we want to keep genes that are selected by both masks but give opposite correlation signs to both masks. The idea behind these experiments is that if a gene is highly positively (or negatively) correlated with both the LN metastases mask and the Bone metastases mask, then it is likely to be a common gene in individuals and not associated with cancer. So this additional gene selection processes will attempt to filter out these genes.\\

The results show that genes that are selected by both the LN and Bone masks but give opposite correlation signs (which we named ``LN-Bone strong genes") produce good results. Next, we consider only the intersecting genes of the ``three masks intersect genes" and the ``LN-bone strong genes" with mask selection threshold of 0.2.\\

\begin{figure}[h!]
\centering
\includegraphics[scale=0.6]{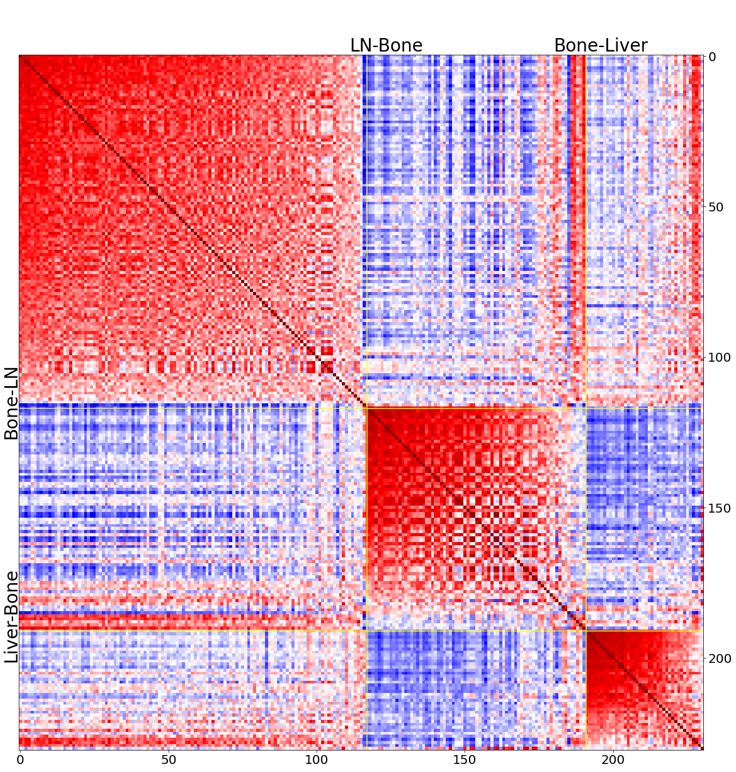}
\caption{\textit{Heatmaps of correlation between any of 231 patients’ selected genes correlation is calculated using rank normalisation, with threshold set to 0.2. There are 133 genes considered which are intersect genes  of  “three masks intersect genes” and “LN-Bone strong genes”}}
\label{fig:133hm}
\end{figure}

Based on results in Fig \ref{fig:133hm}, it is obvious intersecting genes of the “three masks intersect genes” and the “LN-bone strong genes” with mask selection threshold of 0.2 gives a more discriminating heatmap, which is what we would expect to see from the selected genes. This 133 genes set is also very informative and important gene set, so we refer to these genes as SET\_133. As SET\_133 is obtained from SET\_559 with stricter constraints, SET\_133 is a subset of SET\_559. This gene set will be used later in the chapter “Machine Learning classification” and “Gene Co-expression Network and Community Analysis”.\\

\section{Summary}

In this section, we did a though data analysis on the prostate cancer gene expression value dataset, and by a large amount of experiments, we find a effective method to pre-process and pre-select cancer related genes.
In summary, the logic of gene selection process can be expressed in the following way:  

\begin{enumerate}[\hspace{3em}a.]
\item Do a data cleansing to remove obviously meaningless genes.
\item Apply rank normalisation on gene expression dataset
\item Select genes with mask selection method.
\end{enumerate}

In this particular work, the mask selection method can be represented by:

\begin{gather*} 
	    Selected \ gene \ set =  (\lvert C_{Liver}\rvert \geqslant t)\wedge (\lvert C_{LN}\rvert \geqslant t) \wedge (\lvert C_{Bone}\rvert \geqslant t) \wedge (C_{LN}*C_{Bone} < 0)
\end{gather*}

where $t$ is the selection threshold, $t$=0.2,  $C_{LN}$ is the correlation of a gene with the LN metastases mask, $C_{Bone}$ is the correlation of a gene with the Bone metastases mask,  $C_{Liver}$ is the correlation of a gene with the Liver metastases mask. After this process, only 133 out of 14,379 genes are selected.

\chapter{Machine Learning Classification}
One of the aims of this project is to filter out genes related to prostate cancer metastasis and on this basis to classify prostate cancer metastasis secondary sites tumours for diagnostic purposes. Many literature has discussed the use of machine learning methods to classify and predict different cancer types based on differences in gene expression values. However, this project is challenging because essentially the dataset for this work is all prostate cancer cell samples, meaning that they are all cell samples from the same type of cancer. The only difference is that the samples are from different sites of secondary metastasis. This fact suggests that only certain few genes are associated with prostate cancer cell metastasis in these samples, while most genes are not associated with cancer metastasis at all. This requires us to build a sharp and sensitive model that can identify a small number of genes from thousands of genes. This is a challenging task for a machine learning model because the number of genes(features) is much larger than the number of tumour samples(datapoints). When the number of features is greater than the number of datapoints, machine learning models often suffer from the curse-of-dimensionality problem \cite{Bel57} and the over-fitting problem \cite{Tet95}. So to make the machine learning models avoid these problems and have a better performance, we did a gene pre-filter in the previous section. In this section, we use SET\_133 as a start. That is, we only consider 133 genes from 231 samples in this section.\\

\section{XGBoost Model}
There are also many machine learning models available for classification and prediction tasks. However, we decided to use the supervised machine learning algorithm XGBoost (Extreme Gradient Boosting), which is based on decision trees. Although a complex model like a neural network model is powerful and can discover the underlying patterns behind the data, it suffers from an interpretability problem. It is like a ``black box" that produces a single result when we throw data to it, but we have no way of knowing how it makes decisions and which genes play a key role in the decision. XGBoost provides a way to assess the importance of features. It supports giving each feature a score from 0 to 1. We can determine which features play an important role in decision making and which do not by ranking all features according to their importance scores. This advantage is particularly important for this project because, in addition to wanting to know which prostate cancer secondary site our model is classifying and predicting, we also want to know why this is the case and which genes are most important in the decision.\\

\section{Data Balancing}
\begin{figure}[h!]
\centering
\includegraphics[scale=0.35]{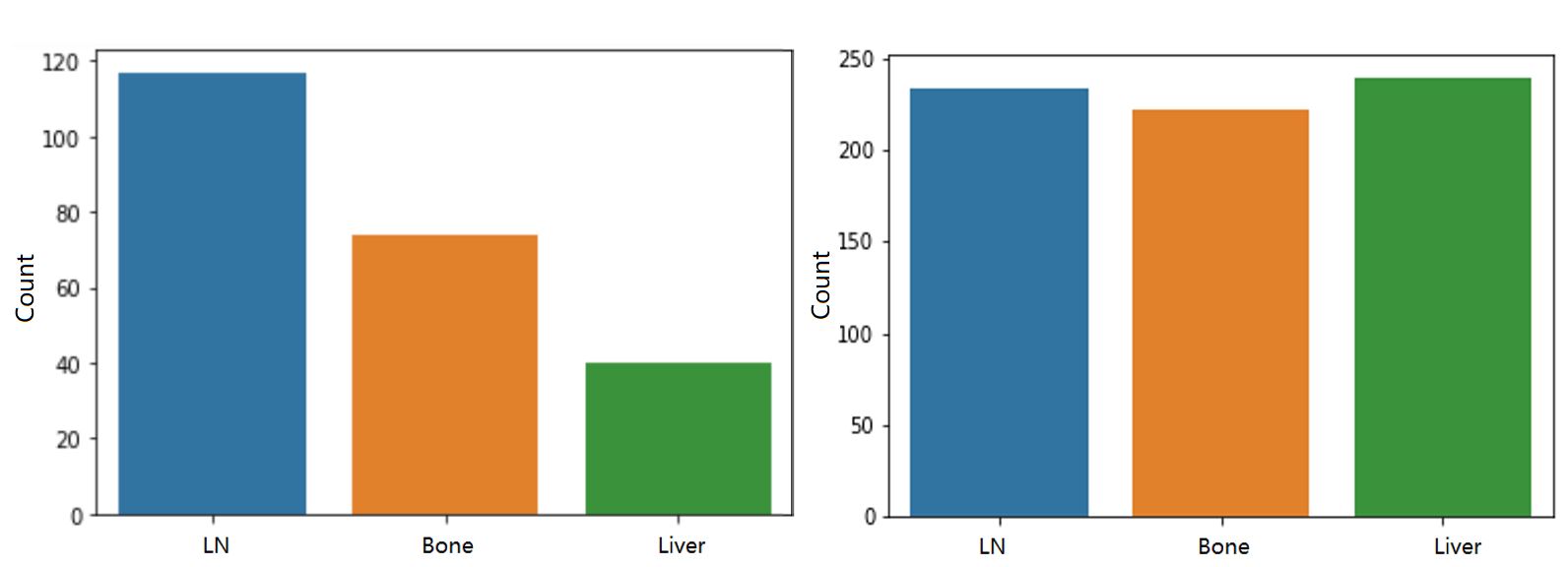}
\caption{\textit{Secondary site types distribution in the original dataset (Left) and Over-sampling balanced dataset (right).}}
\label{fig:databal}
\end{figure}

The bar chart on the left in Figure \ref{fig:databal} shows the distribution of prostate cancer secondary site types in the original dataset, and we note that there is a very unbalanced sample size between the three secondary site types. The unevenness in the number of sample types is likely to have a negative impact on the performance of the model and the fairness of the model being evaluated. As the quality of the dataset is also a key factor in training a good model, we then did a over-sampling data balancing on the original SET\_133 dataset.\\

Firstly, the dataset was randomly divided into 10 sets with the same (or near the same) amount of data and ensured that each set still maintained the same distribution of the three secondary site types. A typical set contains 12 LN, 7 Bone and 4 Liver patient samples. Secondly, within each set, each set is rebalanced by replicating the data for all samples for each secondary site one or more times, so that the number of samples for each site type is approximately equal to the least common multiple of the three site types. Here, we replicate the LN sample data once, Bone twice and Liver five times, at which point there are approximately 24 LNs, 21 Bones and 24 Livers patient samples in a set. When doing 10-fold cross-validation, one set of data at a time is selected as the validation set and the other nine as the training set. This was to ensure that both a sample and its replicate were in the same training or validation set. If a sample appears in the training set and its copy appears in the validation dataset, the validation results will be corrupted. \\  

\section{Key Genes Selection }
\begin{figure}[h!]
\centering
\includegraphics[scale=0.45]{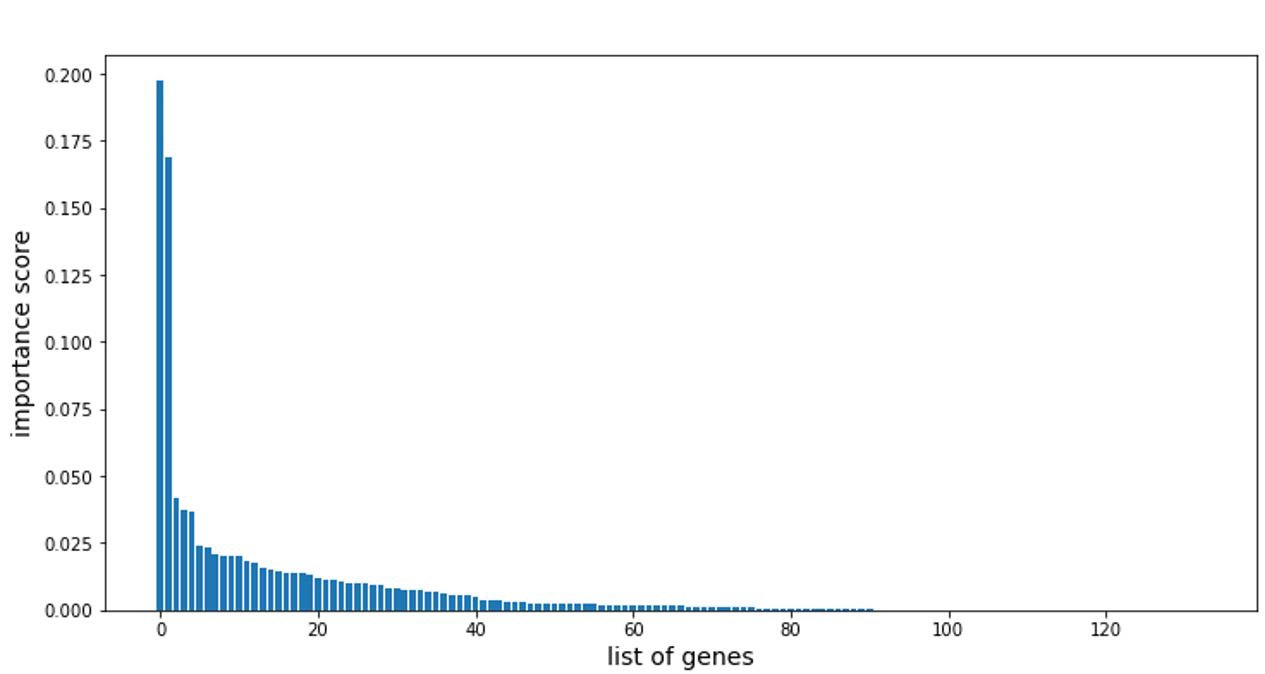}
\caption{\textit{Feature importance of all 133 features  in the trained XGBoost model.}}
\label{fig:133importance}
\end{figure}

In the first step, we trained the XGBoost model directly on the unbalanced SET\_133 dataset and used 10-fold cross-validation to train and validate the model. We obtained a result of 92.21\% overall accuracy. In terms of classification accuracy, the model performed well, but we found that not all 133 genes played a role in the predictions. Figure \ref{fig:133importance} shows that only about 40 genes played a significant role and, of these, two were particularly important. However, most genes were not important and even about 40 genes had an importance score of 0, meaning that these genes did not affect prediction at all.

Based on this observation, we are going to remove the unimportant genes. The general idea is to train a model, remove the least important genes each time and retrain the model to evaluate its accuracy. By removing the trailing genes step by step and observing the corresponding model accuracy, we find the set of genes used when the accuracy is highest.\\

We performed the following detailed steps to filter the significant genes.\\

\begin{enumerate}[\hspace{3em}a.]
\item Train and validate the model by performing a 10-fold cross-validation with the original SET\_133 genes. The validation accuracy and feature importance scores in the XGBoost model are then calculated.
\item Remove the least important genes based on the feature importance scores.
\item With the remaining genes, train and validate the model using 10-fold cross-validation. Calculate the validation accuracy and feature importance score of the XGBoost model. Repeat 10-fold cross-validation 10 times and take the average accuracy value.
\item Cycle through steps b to c until only 3 genes remain.
\item Plot a graph showing the accuracy of model validation versus the number of tail importance genes discarded.
\item Then apply an oversampling technique to balance the dataset, also doing the above steps.
\end{enumerate}

\begin{figure}[h!]
\centering
\includegraphics[scale=0.5]{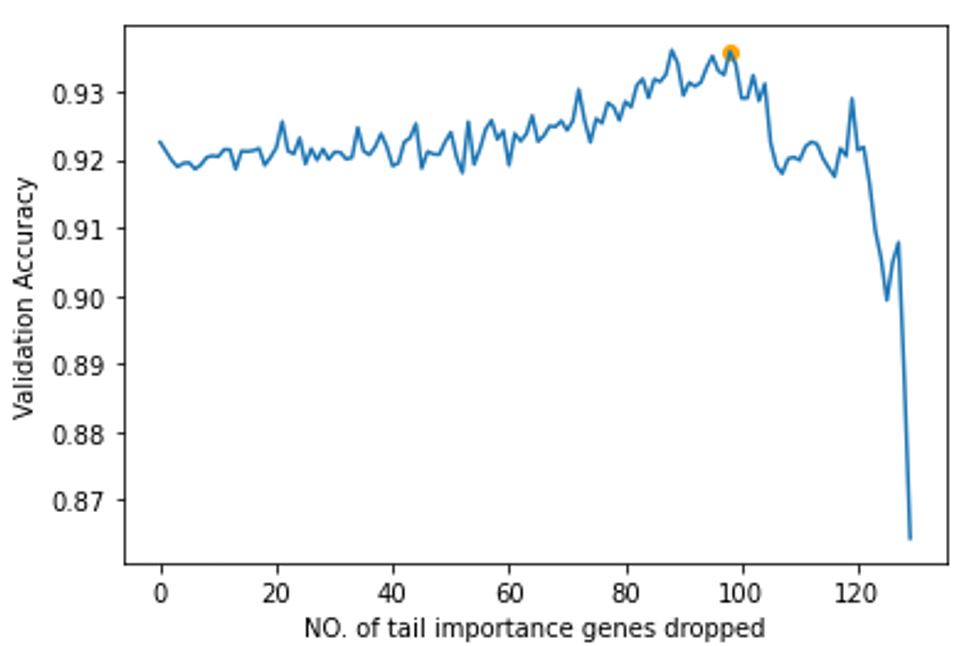}
\caption{\textit{XGBoost model 10-fold cross-validation accuracy against number of tail importance genes dropped with original imbalanced SET\_133 dataset. }}
\label{fig:oridrop}
\end{figure}

\begin{figure}[h!]
\centering
\includegraphics[scale=0.5]{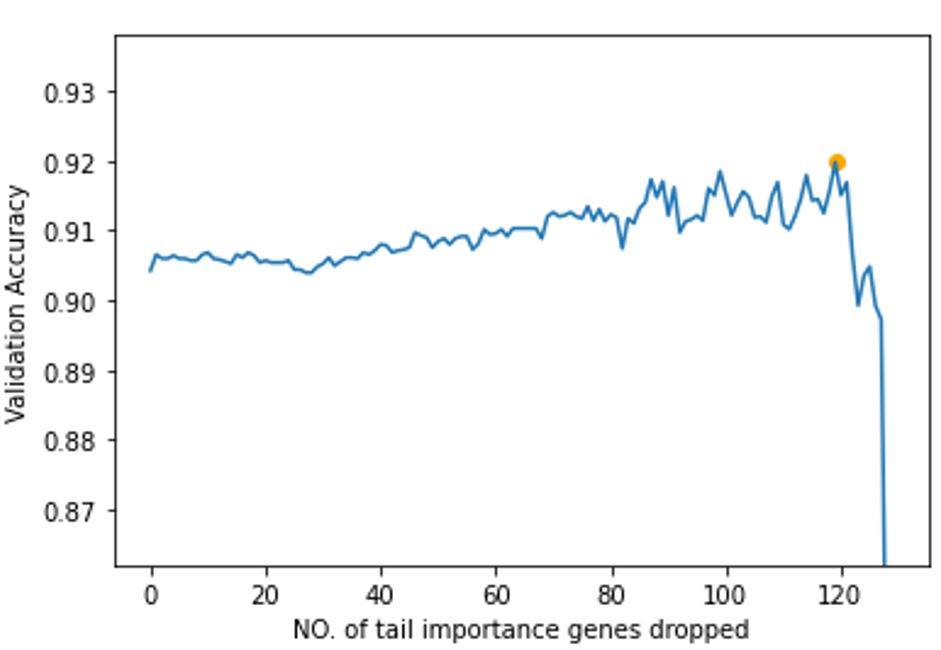}
\caption{\textit{XGBoost model 10-fold cross-validation accuracy against number of tail importance genes dropped with the over-sampling balanced SET dataset. }}
\label{fig:baldrop}
\end{figure}

From the Figure \ref{fig:oridrop}, we can see that for the original unbalanced SET\_133 dataset, the model can still have a high average cross-validation accuracy of 93.58\%, even after dropping 99 tail significant genes, i.e. only 34 genes remain. (Orange dots in Figure \ref{fig:oridrop}.) We named the set of 34 genes SET\_34. From the Figure \ref{fig:baldrop}, we can see that the highest accuracy of 91.97\% was obtained with only 13 genes remaining in the case of over-sampling the balanced SET dataset. (Orange dots in the Figure \ref{fig:baldrop}) We named the set of 13 genes as SET\_13.\\

Comparing these two plots in Figure \ref{fig:oridrop} and \ref{fig:baldrop}, we find that the validation accuracy of the model with the balanced over dataset is generally lower than that of the model with the unbalanced dataset. This does not necessarily mean that the model using balanced data has poorer performance. Instead, as unbalanced data potentially biases the validation result, the model using balanced data may more closely represent the actual performance of the model.\\

Interestingly, the line trend in both Figure \ref{fig:oridrop} and \ref{fig:baldrop} shows that the validation accuracy in general continues to improve when unimportant genes are eliminated. This contradicts our initial assumption that excluding genes would imply a loss of information. However, this result may suggest that information about prostate cancer metastasis may be encoded in only a very small number of genes.\\

We also noticed that the accuracy line was a fluctuating increase, rather than a smooth increase when the least important genes were removed each time. This could be explained by the fact that genes may have co-expression patterns, and if we remove a least important co-expressed gene cluster each time, we may get a smoother accuracy graph. However, at present we do not know which genes are in the same co-expression cluster without the further help of our biological knowledge. However, the gene co-expression network analysis may be of some help in this regard.\\

Another important discovery we made was that for the original dataset, while the best results were obtained when 34 genes were retained, there was also a significant local maximum when 14 genes were retained. Looking further into the details of these genes we found that these the SET\_13 genes are all included in these 14 genes, which also implies that the genes in SET\_13 are important.\\

The top important 13 genes in the SET\_13 are listed below:
PPP1R16A, CXXC1, MAGED2, LRRC66, ERGIC3, GARS, RPL23, RAP1GAP2, IQCE, UBE2L3, UBE2Z, CARD16, PTS.\\

\begin{figure}[h!]
\centering
\includegraphics[scale=0.4]{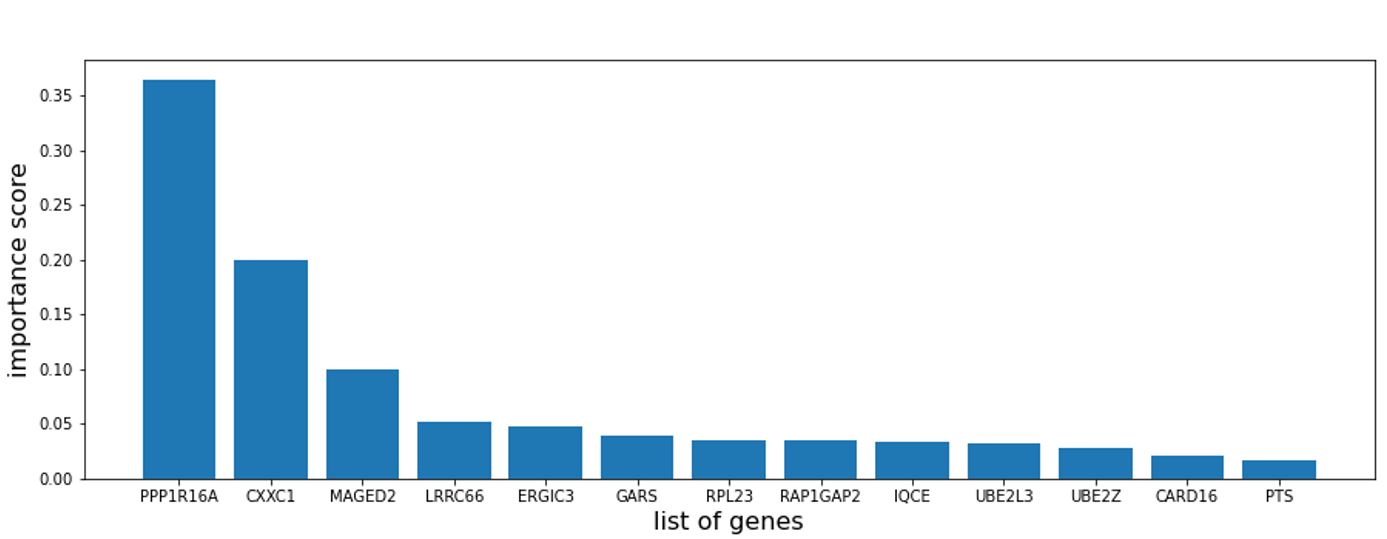}
\caption{\textit{The 13 genes in SET\_13 which are considered as very important for prostate cancer metastasis. Genes are ordered in descending order of importance score in the XGBoost classification model.}}
\label{fig:top13}
\end{figure}

\section{Classification Result Evaluation}

\begin{figure}[h!]
\centering
\includegraphics[scale=0.55]{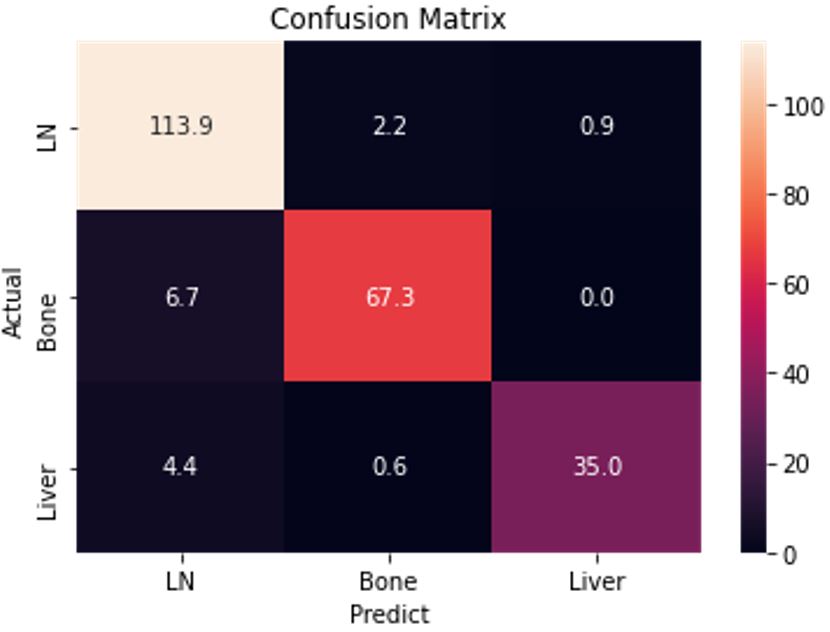}
\caption{\textit{Confusion matrix for the 10-fold cross validation set predictions with original imbalanced dataset.}}
\label{fig:oriconfnx}
\end{figure}

\begin{figure}[h!]
\centering
\includegraphics[scale=0.55]{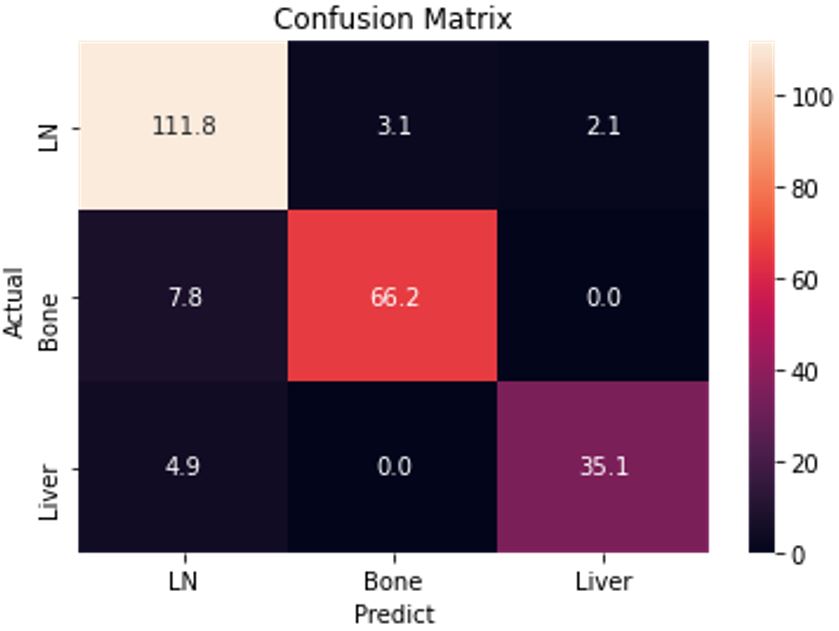}
\caption{\textit{Confusion matrix for the 10-fold cross validation set predictions with the over-sampling balanced dataset.}}
\label{fig:balconfnx}
\end{figure}

\begin{table}[]
\centering
\begin{tabular}{llllllll}
\cline{1-4}
\multicolumn{1}{|l|}{} & \multicolumn{1}{l|}{\textbf{precision}} & \multicolumn{1}{l|}{\textbf{recall}} & \multicolumn{1}{l|}{\textbf{f1-score}} &  &  &  &  \\ \cline{1-4}
\multicolumn{1}{|l|}{\textbf{LN}} & \multicolumn{1}{l|}{0.91} & \multicolumn{1}{l|}{0.97} & \multicolumn{1}{l|}{0.94} &  &  &  &  \\ \cline{1-4}
\multicolumn{1}{|l|}{\textbf{Bone}} & \multicolumn{1}{l|}{0.96} & \multicolumn{1}{l|}{0.91} & \multicolumn{1}{l|}{0.93} &  &  &  &  \\ \cline{1-4}
\multicolumn{1}{|l|}{\textbf{Liver}} & \multicolumn{1}{l|}{0.97} & \multicolumn{1}{l|}{0.88} & \multicolumn{1}{l|}{0.92} &  &  &  &  \\ \cline{1-4}
 &  &  &  &  &  &  & 
\end{tabular}
\caption{\textit{Precision, recall and F1 score of the XGBoost model prediction result with the original imbalanced dataset.}}
\label{tab:oritable}
\end{table}

\begin{table}[]
\centering
\begin{tabular}{llllllll}
\cline{1-4}
\multicolumn{1}{|l|}{} & \multicolumn{1}{l|}{\textbf{precision}} & \multicolumn{1}{l|}{\textbf{recall}} & \multicolumn{1}{l|}{\textbf{f1-score}} &  &  &  &  \\ \cline{1-4}
\multicolumn{1}{|l|}{\textbf{LN}} & \multicolumn{1}{l|}{0.90} & \multicolumn{1}{l|}{0.96} & \multicolumn{1}{l|}{0.93} &  &  &  &  \\ \cline{1-4}
\multicolumn{1}{|l|}{\textbf{Bone}} & \multicolumn{1}{l|}{0.96} & \multicolumn{1}{l|}{0.89} & \multicolumn{1}{l|}{0.92} &  &  &  &  \\ \cline{1-4}
\multicolumn{1}{|l|}{\textbf{Liver}} & \multicolumn{1}{l|}{0.94} & \multicolumn{1}{l|}{0.88} & \multicolumn{1}{l|}{0.91} &  &  &  &  \\ \cline{1-4}
 &  &  &  &  &  &  & 
\end{tabular}
\caption{\textit{Precision, recall and F1 score of the XGBoost model prediction result with the over-sampling balanced dataset.}}
\label{tab:baltable}
\end{table}

The performance of this ML prediction model is impressive, with an overall prediction accuracy of 93.58\% and 91.97\% on the validation dataset using an unbalanced dataset and an over-sampled balanced dataset respectively.\\

As we can see from the confusion matrix and the precision, recall and F1 score tables, the model can easily classify LN samples correctly, but they also have a slight tendency to classify some bone and liver samples as LN samples. The high precision of these models in classifying bone and liver suggests that few other samples would be misclassified as bone and liver.\\

These results may indicate that we have selected a very important set of genes for the very accurate classification of secondary sites of prostate cancer. However, the relatively high precision and relatively low recall of the LN classification, and the relatively high precision and the relatively low recall of the bone and liver classification may indicate the following:

\begin{enumerate}[\hspace{3em}a.]
\item The selected genes set may have included most of for LN metastases related genes, but still missed some. This may explain why the model can correctly classify LN samples but also misclassify other samples as LN samples. 

\item The genes set we selected may have contained more than enough bone and liver  metastases related genes. And the model may have used these excess genes for LN, bone and liver classification, which resulted in few other samples being misclassified as bone and liver, but some bone and liver samples being misclassified as LN samples.\\

\end{enumerate}

\chapter{Gene Co-expression Network and Community Analysis}
\section{Gene Co-expression Network (GCN)}
\subsection{Intuition of Gene Co-expression Network }
As described in the background section, a set of co-expressed genes are involved in the synthesis of a functionally complex protein, and finding a set of genes associated with prostate cancer metastasis would provide a deeper understanding of this cancer. Pure bioanalysis to find co-expressed genes is often difficult and requires a great deal of work. It is fortunate that the way genes are co-expressed is intuitively consistent with the concept of network science. If genes and their co-expression information can be represented as a topological network, advanced network analysis techniques such as community detection can be applied to extract further information and knowledge about genes. Therefore, the concept of a gene co-expression network (GCN) was proposed \cite{Stu03} to represent gene co-expression information and relationships. The construction of a GCN is concept-wise straightforward: nodes represent genes, and if there is an obvious co-expression relationship between a pair of nodes, then an undirected edge is used to connect them.\\

In this project, we have argued in ``Background" section that the Pearson correlation coefficient are often used to measure the relationship between two gene co-expressions. Therefore, gene co-expression networks can be constructed by correlation coefficients between genes.\\

\subsection{Dataset}
For the GCN analysis, we would like to consider a relatively large gene set, as a small gene set would provide limited information. In this chapter, we will use SET\_559 as a start to construct the GCN. In the network analysis, we will also focus on the distribution of SET\_133, SET\_34 and SET\_13 genes in a GCN. We are interested in these gene sets because they are very important gene sets, and any obvious distributive pattern of these critical genes may reveal important underlying knowledge. Note that SET\_13 is a subset of SET\_34, SET\_34 is a subset of SET\_133, and SET\_133 is a subset of SET\_559. We define the set difference notation as follows:\\

\begin{gather*} 
	    A - B = \{x: x \in A and x \notin B\} for \ set \ A \ and \ set \ B
\end{gather*}

For example, SET\_34 – SET\_13 represents a set of genes which in SET\_34 but not in SET\_13.

\subsection{Network Construction }

We use the 559 nodes in the network to represent each gene in the SET\_559. To analyse the GCNs under different prostate cancer secondary site types, we will construct four different GCNs, one for all 231 samples and the other three for each prostate cancer secondary site type. We will refer to the GCN for prostate cancer cells at the LN site as the LN GCN, the GCN for prostate cancer cells at the Bone site as the Bone GCN, the GCN for prostate cancer cells at the Liver site as the Liver GCN, and the GCN for all sample cells as the All-patient GCN. Since the All-patient GCN may contain an ensemble information from all LN , Bone and Liver samples, it is difficult to get clear information by analysing this network, so we just use this network as a reference. Our work will focus on analysing and comparing the LN, Bone and Liver GCNs.\\

Using the Bone GCN as an example, we build the network by the following steps: \\

\begin{enumerate}[\hspace{3em}a.]
\item For each of the 559 nodes (genes), add a weighted edge between any two nodes, where the weight value is the absolute value of the correlation coefficient between these two genes. It is important to note that when we construct the Bone GCN, we should only consider the 74 Bones samples. Therefore, the correlation between the two genes is only calculated based on these 74 values. After this step, we will have a complete weighted network, which means that there is an edge between any two nodes and all the edges are weighted.  

\item Set thresholds from 0.4 to 0.9 with step 0.02. For each threshold, construct the corresponding gene network by removing all edges which weight less than this threshold, and keep all edges which weight greater than or equal to this threshold. Ignore all weights of all edges, then we will obtain an unweighted Bone GCN.

\item Run non-overlapping community detection algorithm on each gene network, then calculate the network modularity \cite{New06}.

\item Among all the GCNs constructed with candidate thresholds, keep the GCN with the highest modularity value.
\end{enumerate}

The LN and Liver GCNs could also be constructed by the above steps, by considering only the Bone and Liver samples when calculating the correlation values. \\

\begin{figure}[h!]
\centering
\includegraphics[scale=0.58]{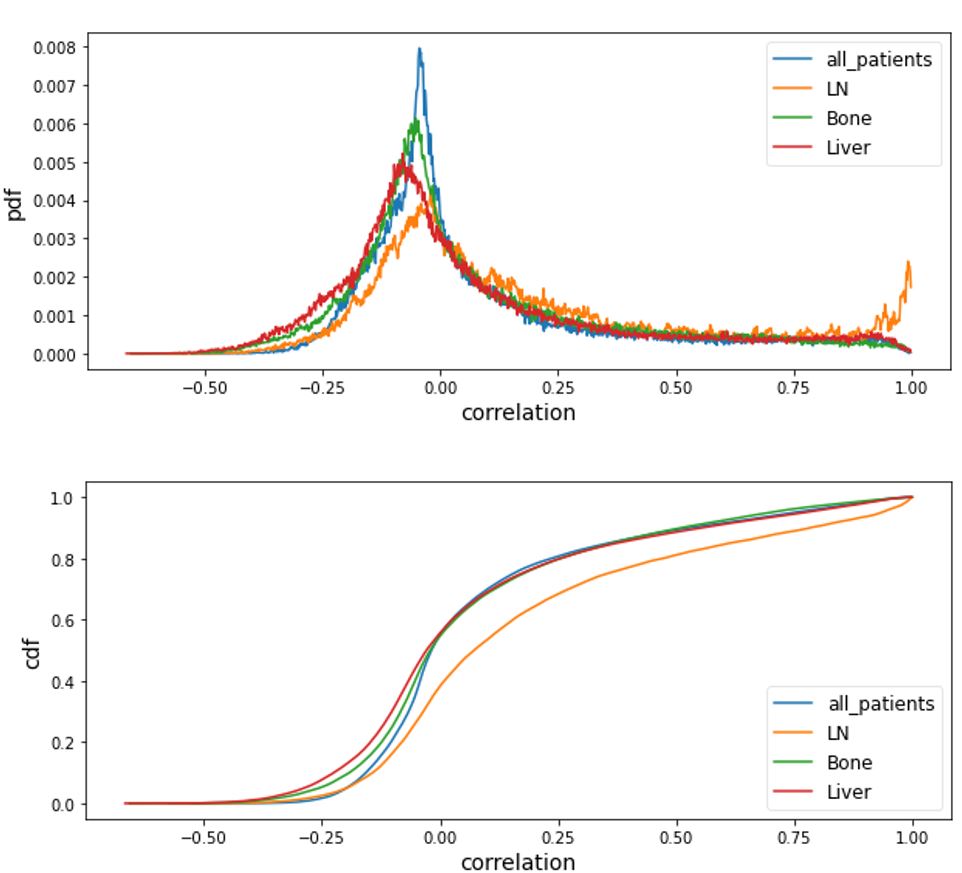}
\caption{\textit{Probability distribution function (PDF) and Cumulative distribution function (CDF) of Pearson correlation between 559 genes for each secondary site patients. Up: PDF, Down: CDF}}
\label{fig:corrpdfcdf}
\end{figure}

As can be seen from the probability distribution function of Pearson correlations for samples at each secondary site in Figure \ref{fig:corrpdfcdf}, most of the edges have low weight values in all three GCNs, which is expected because not most genes are co-expressed with each other. We may be able to construct sparse GCNs, which may make it relatively easy to analyse valuable information. However, we also noted that the LN GCN may differ from the other two GCNs in that there are a relatively large number of genes with very high correlations, suggesting that the LN GCN is denser than the other two GCNs, i.e. the average node degree of the LN may be significantly greater. This could cause difficulties for our later comparisons with these GCNs.\\

Much of the literature uses weighted GCN for analysis, but co-expression is binary in the sense of biology. This means that two genes are either co-expressed or not co-expressed in the production of a certain protein, with no intermediate situation. For a clear explanation, consider this situation. A gene, 'A', has a correlation of 0.01 with gene 'B', 0.3 with gene 'C', 0.7 with gene 'D' and 0.9 with gene 'E'. It is possible that 'A' has no co-expression relationship with either 'B' or 'C', while it has co-expression relationships with both 'C' and 'D'. Even if the correlation between 'A' and 'C' is significantly greater than the correlation with 'B', it does not necessarily mean that 'A' is more co-expressed with 'C' than with 'B'. Similarly, even if the strong relationship between "A" and "D" is significantly smaller than the relationship with "E", it does not necessarily mean that "A" has less co-expression with "D" than with "E". We therefore wish to reduce the impact of this misleading information by setting a threshold and converting the weighted network to unweighted.

\section{Community Detection}
For genes with strong co-expression relationships, strong community patterns are expected in GCN. GCN may have multiple distinct community structures, with each gene community representing a biological cluster of gene co-expressions that are responsible for the synthesis of a number of specific proteins. Therefore, we can perform a community detection on GCN to identify clusters of co-expressed genes. It is likely that some genes are involved in the synthesis of multiple proteins, which means that there may be overlapping communities in the GCN. However, as this work is an initial study on genetic analysis on metastatic prostate cancer with limited time, we will focus only on non-overlapping community detection to simplify our work.\\

In network science, a network property called modularity is a measure of the strength of a network's division into communities.  Networks with a high degree of modularity tend to have a distinct community structure. Many non-overlapping community detection algorithms have been designed to divide networks into communities by optimising network modularity. The algorithm used in this work is the default non-overlapping community detection algorithm implemented in Gephi, which is also a typical modularity-based algorithm \cite{Vin08}.\\

Before starting the network analysis of the LN, Bone and Liver GCNs, we first performed community detection on the All-patients network to see if there were any obvious communities. The All-patients GCN was constructed from an edge threshold of 0.5, at which point modularity reached a maximum of 0.52. In this network there were 16,704 edges and 502 non-isolated nodes.  However, there are 5 tiny groups disconnected from the giant components of the network, so we also treat them as unimportant genes. Thus, there are 492 genes in the network forming 6 distinguishable communities. Since we consider that the GCNs of all patients may contain information about all of LN, Bone and Liver GCNs, we then consider only these 492 genes further in LN, Bone and Liver GCN construction.\\

In the previous section, we noted that the LN GCN may be somewhat different from the Bone and Liver GCNs. So, we first construct the Bone and Liver GCNs, and then refer to the network characteristics of the Bone and Liver GCNs before constructing the LN GCN. When necessary, we may make some trade-offs on the LN GCN to make the three networks comparable.\\

\subsection{Bone GCN}
The Bone GCN was constructed from an edge threshold of 0.62, at which point modularity reached a maximum of 0.54. In this network, there are 10,554 edges and 431 nodes in the giant component, which has an average node degree of 48.9. The network is divided into 13 communities and details of the distribution of important genes in these communities are shown in Table \ref{tab:Bonecom}. The visualisation of the Bone GCN can be found in Table \ref{tab:colcomp} under section “7.2.4 Community Comparison Among Different GCNs”, and the "Colour" column in table \ref{tab:Bonecom} corresponds to the community colour scheme of the network visualisation in that section. For all genes in each community, the number of genes appearing in SET\_13 is placed under the "SET\_13" column, the number of gene appearing in SET\_34 but not in SET\_13 is placed under "SET\_34 - SET\_13", and so on.

\begin{table}[h!]
\begin{tabular}{|l|l|l|l|l|l|l|}
\hline
\multicolumn{2}{|l|}{\textbf{Community}} &  &  &  &  &  \\ \cline{1-2}
\textbf{Colour} & \textbf{Ref.} & \multirow{-2}{*}{\textbf{\begin{tabular}[c]{@{}l@{}}Number \\ of nodes\end{tabular}}} & \multirow{-2}{*}{\textbf{SET\_13 \ }} & \multirow{-2}{*}{\textbf{\begin{tabular}[c]{@{}l@{}}SET\_34\\ - SET\_13\end{tabular}}} & \multirow{-2}{*}{\textbf{\begin{tabular}[c]{@{}l@{}}SET\_133\\ - SET\_34\end{tabular}}} & \multirow{-2}{*}{\textbf{\begin{tabular}[c]{@{}l@{}}SET\_133\\ - SET\_559\end{tabular}}} \\ \hline
{\color[HTML]{6434FC} Violet} & 1 & 154 & 5 & 8 & 42 & 99 \\ \hline
{\color[HTML]{32CB00} Green} & 2 & 96 & 0 & 1 & 3 & 92 \\ \hline
{\color[HTML]{34CDF9} Light blue} & 3 & 67 & 5 & 5 & 23 & 34 \\ \hline
{\color[HTML]{F8A102} Orange} & 4 & 25 & 0 & 1 & 5 & 19 \\ \hline
{\color[HTML]{F8FF00} Yellow} & 5 & 23 & 2 & 1 & 3 & 17 \\ \hline
{\color[HTML]{FE0000} Red} & 6 & 14 & 1 & 1 & 2 & 10 \\ \hline
{\color[HTML]{3531FF} Deep blue} & 7 & 10 & 0 & 0 & 1 & 9 \\ \hline
{\color[HTML]{9B9B9B} Gray} & 8 & 9 & 0 & 0 & 1 & 8 \\ \hline
{\color[HTML]{9B9B9B} Gray} & 9 & 8 & 0 & 0 & 4 & 4 \\ \hline
{\color[HTML]{9B9B9B} Gray} & 10 & 8 & 0 & 1 & 0 & 7 \\ \hline
{\color[HTML]{9B9B9B} Gray} & 11 & 8 & 0 & 0 & 0 & 8 \\ \hline
{\color[HTML]{9B9B9B} Gray} & 12 & 7 & 0 & 0 & 1 & 6 \\ \hline
{\color[HTML]{9B9B9B} Gray} & 13 & 2 & 0 & 0 & 1 & 1 \\ \hline
\multicolumn{2}{|l|}{Total} & \begin{tabular}[c]{@{}l@{}}431\\ (Giant\\ component)\end{tabular} & 13 & 18 & 86 & 314 \\ \hline
\end{tabular}
\caption{\textit{Bone GCN community details}}
\label{tab:Bonecom}
\end{table}

\subsection{Liver GCN}
The Liver GCN was constructed from an edge threshold of 0.54, at which point modularity reached a maximum of 0.33. In this network, there are 15,748 edges and 477 nodes in the giant component, which has an average node degree of 66.4. The network is divided into 6 communities and details of the distribution of important genes in these communities are shown in Table \ref{tab:Livercom}.

\begin{table}[h!]
\begin{tabular}{|l|l|l|l|l|l|l|}
\hline
\multicolumn{2}{|l|}{\textbf{Community}} &  &  &  &  &  \\ \cline{1-2}
\textbf{Colour} & \textbf{Ref.} & \multirow{-2}{*}{\textbf{\begin{tabular}[c]{@{}l@{}}Number \\ of nodes\end{tabular}}} & \multirow{-2}{*}{\textbf{SET\_13 \ }} & \multirow{-2}{*}{\textbf{\begin{tabular}[c]{@{}l@{}}SET\_34\\ - SET\_13\end{tabular}}} & \multirow{-2}{*}{\textbf{\begin{tabular}[c]{@{}l@{}}SET\_133\\ - SET\_34\end{tabular}}} & \multirow{-2}{*}{\textbf{\begin{tabular}[c]{@{}l@{}}SET\_133\\ - SET\_559\end{tabular}}} \\ \hline
{\color[HTML]{6434FC} Violet} & 1 & 171 & 4 & 3 & 21 & 143 \\ \hline
{\color[HTML]{32CB00} Green} & 2 & 164 & 8 & 8 & 33 & 115 \\ \hline
{\color[HTML]{34CDF9} Light blue} & 3 & 124 & 1 & 7 & 33 & 83 \\ \hline
{\color[HTML]{F8A102} Orange} & 4 & 25 & 0 & 0 & 3 & 22 \\ \hline
{\color[HTML]{F8FF00} Yellow} & 5 & 2 & 0 & 0 & 1 & 1 \\ \hline
{\color[HTML]{FE0000} Red} & 6 & 2 & 0 & 0 & 1 & 1 \\ \hline
\multicolumn{2}{|l|}{Total} & \begin{tabular}[c]{@{}l@{}}474\\ (Giant \\ component)\end{tabular} & 13 & 18 & 94 & 349 \\ \hline
\end{tabular}
\caption{\textit{Liver GCN community details}}
\label{tab:Livercom}
\end{table}

\subsection{LN GCN}
By the GCN construction procedure in section ``7.1.3 Network Construction", the LN GCN was constructed from an edge threshold of 0.24, at which point modularity reached a maximum of 0.19. In this network, there are 50,477 edges and 492 nodes in the giant component, which has an average node degree of 205.2.  This network is very different from the Bone and Liver GCN in terms of network size, average node degree and modularity. The low threshold allows a large number of edges to be preserved, making the network very dense. We also noted that if we use larger thresholds, the number of nodes in the giant component becomes smaller, but the average node degree of the network remains high and the modularity of the network keeps decreasing. In order to make the LN GCN comparable to the Bone and Liver GCNs, we made a trade-off here by customising the threshold so that the size of the giant component of the LN GCN is similar to that of the Bone and Liver GCNs. 

The LN GCN was then constructed from an edge threshold of 0.56, with 25,685 edges, 424 nodes in the giant component, which has an average node degree of 118.9, eight communities and a modularity of 0.10.

\begin{table}[]
\begin{tabular}{|l|l|l|l|l|l|l|}
\hline
\multicolumn{2}{|l|}{\textbf{Community}} &  &  &  &  &  \\ \cline{1-2}
\textbf{Colour} & \textbf{Ref.} & \multirow{-2}{*}{\textbf{\begin{tabular}[c]{@{}l@{}}Number \\ of nodes\end{tabular}}} & \multirow{-2}{*}{\textbf{SET\_13 \ }} & \multirow{-2}{*}{\textbf{\begin{tabular}[c]{@{}l@{}}SET\_34\\ - SET\_13\end{tabular}}} & \multirow{-2}{*}{\textbf{\begin{tabular}[c]{@{}l@{}}SET\_133\\ - SET\_34\end{tabular}}} & \multirow{-2}{*}{\textbf{\begin{tabular}[c]{@{}l@{}}SET\_133\\ - SET\_559\end{tabular}}} \\ \hline
{\color[HTML]{6434FC} violet} & 1 & 212 & 5 & 6 & 41 & 160 \\ \hline
{\color[HTML]{32CB00} green} & 2 & 115 & 3 & 6 & 26 & 80 \\ \hline
{\color[HTML]{34CDF9} Light blue} & 3 & 40 & 3 & 3 & 6 & 28 \\ \hline
{\color[HTML]{F8A102} Orange} & 4 & 25 & 0 & 0 & 3 & 22 \\ \hline
{\color[HTML]{F8FF00} Yellow} & 5 & 10 & 0 & 0 & 1 & 9 \\ \hline
{\color[HTML]{FE0000} Red} & 6 & 9 & 1 & 1 & 0 & 7 \\ \hline
{\color[HTML]{3531FF} Deep Blue} & 7 & 8 & 0 & 0 & 1 & 7 \\ \hline
{\color[HTML]{FFCCC9} Pink} & 8 & 5 & 0 & 0 & 0 & 5 \\ \hline
\multicolumn{2}{|l|}{Total} & \cellcolor[HTML]{FFFFFF}\begin{tabular}[c]{@{}l@{}}424\\ (giant \\ component)\end{tabular} & 12 & 16 & 78 & 318 \\ \hline
\end{tabular}
\caption{\textit{LN GCN community details}}
\label{tab:LNrcom}
\end{table}

\subsection{Community Comparison Among Different GCNs }

\begin{table}[]
\begin{tabular}{|r|l|l|l|l|}
\hline
\multicolumn{1}{|l|}{\textbf{\begin{tabular}[c]{@{}l@{}}Gene \\ Co-expression\\ Networks\end{tabular}}} & \textbf{\begin{tabular}[c]{@{}l@{}}All-patients   \\ GCN\end{tabular}} & \textbf{LN GCN} & \textbf{Bone GCN} & \textbf{Liver GCN} \\ \hline
\begin{tabular}[c]{@{}r@{}}Number of \\  samples\end{tabular} & 231 & 117 & 74 & 40 \\ \hline
\begin{tabular}[c]{@{}r@{}}Number of   \\ genes under study\end{tabular} & 559 & \cellcolor[HTML]{FFF2CC}492 & \cellcolor[HTML]{FFF2CC}492 & \cellcolor[HTML]{FFF2CC}492 \\ \hline
\begin{tabular}[c]{@{}r@{}}Correlation\\ Thresshold\end{tabular} & 0.50 & 0.56 & 0.62 & 0.54 \\ \hline

\begin{tabular}[c]{@{}r@{}}Number of \\ non-isolated nodes\end{tabular} & 502 & 432 & 443 & 477 \\ \hline

\begin{tabular}[c]{@{}r@{}}{\color[HTML]{FFFFFF} .}\\   Number of links\end{tabular}  & 16,704 & 25,685 & 10,554 & 15,748 \\ \hline

\begin{tabular}[c]{@{}r@{}}Average \\ node degree\end{tabular} & 66.5 & 118.9 & 47.6 & 66.0 \\ \hline
\begin{tabular}[c]{@{}r@{}}{\color[HTML]{FFFFFF} .} \\ Modularity \end{tabular}& 0.52 & 0.10 & 0.54 & 0.33 \\ \hline
\multicolumn{5}{|c|}{\textbf{Giant Component}} \\ \hline
\begin{tabular}[c]{@{}r@{}}Giant Component\\  size\end{tabular} & \cellcolor[HTML]{FFF2CC}492 & 424 & 431 & 474 \\ \hline
\begin{tabular}[c]{@{}r@{}}{\color[HTML]{FFFFFF} .} \\ Number of  links \end{tabular} & 16,699 & 25,681 & 10,548 & 15,746 \\ \hline
\begin{tabular}[c]{@{}r@{}}Average \\  node degree\end{tabular} & 67.9 & 121.1 & 48.9 & 66.4 \\ \hline
\begin{tabular}[c]{@{}r@{}}{\color[HTML]{FFFFFF} .} \\ Modularity \end{tabular}& 0.52 & 0.10 & 0.54 & 0.33 \\ \hline
\multicolumn{5}{|c|}{\textbf{Community}} \\ \hline
\begin{tabular}[c]{@{}r@{}}Number of\\ communities\end{tabular} & 6 & 8 & 13 & 6 \\ \hline
\begin{tabular}[c]{@{}r@{}}{\color[HTML]{6434FC} 1st community size}\\    (list of   key genes)\end{tabular} & \begin{tabular}[c]{@{}l@{}}154\\ (2,3,6)\end{tabular} & \begin{tabular}[c]{@{}l@{}}212\\ (2,6,8,11,12)\end{tabular} & \begin{tabular}[c]{@{}l@{}}154\\ (0,2,3,11,12)\end{tabular} & \begin{tabular}[c]{@{}l@{}}171\\ (2,3,6,11)\end{tabular} \\ \hline
\begin{tabular}[c]{@{}r@{}} {\color[HTML]{32CB00}2nd community size}\\   (list of   key genes)\end{tabular} & \begin{tabular}[c]{@{}l@{}}130\\ (0,11,12)\end{tabular} & \begin{tabular}[c]{@{}l@{}}115\\ (4,5,9)\end{tabular} & 96 & \begin{tabular}[c]{@{}l@{}}164\\ (0,1,5,7,8,9,10,12)\end{tabular} \\ \hline
\begin{tabular}[c]{@{}r@{}}{\color[HTML]{34CDF9}3rd community size}\\   (list of   key genes)\end{tabular} & \begin{tabular}[c]{@{}l@{}}70\\ (7,10)\end{tabular} & \begin{tabular}[c]{@{}l@{}}40\\ (0,7,10)\end{tabular} & \begin{tabular}[c]{@{}l@{}}67\\ (1,5,6,8,10)\end{tabular} & \begin{tabular}[c]{@{}l@{}}124\\ (4)\end{tabular} \\ \hline
\begin{tabular}[c]{@{}r@{}}{\color[HTML]{F8A102}4th community size}\\   (list of   key genes)\end{tabular} & 60 & 25 & 25 & 25 \\ \hline
\begin{tabular}[c]{@{}r@{}}{\color[HTML]{F8FF00}5th community size}\\   (list of   key genes)\end{tabular} & \begin{tabular}[c]{@{}l@{}}51\\ (1,5,8,9)\end{tabular} & 10 & \begin{tabular}[c]{@{}l@{}}23\\ (7,9)\end{tabular} & 2 \\ \hline

\begin{tabular}[c]{@{}r@{}}{\color[HTML]{FE0000}6th community size}\\   (list of   key genes)\end{tabular} & \begin{tabular}[c]{@{}l@{}}27\\ (4)\end{tabular} & \begin{tabular}[c]{@{}l@{}}9\\ (1)\end{tabular} & \begin{tabular}[c]{@{}l@{}}14\\ (4)\end{tabular} & 2 \\ \hline

\begin{tabular}[c]{@{}r@{}}{\color[HTML]{FFFFFF} .} \\ (isolated   key genes) \end{tabular}&  &\begin{tabular}[c]{@{}l@{}} {\color[HTML]{FFFFFF} .} \\ (3)\end{tabular} &  &  \\ \hline
\end{tabular}
\caption{\textit{Summary of the All-patient, LN, Bone and Liver GCN communities. }}
\label{tab:comsum}
\end{table}


\begin{figure}[h!]
\centering
\includegraphics[scale=0.5]{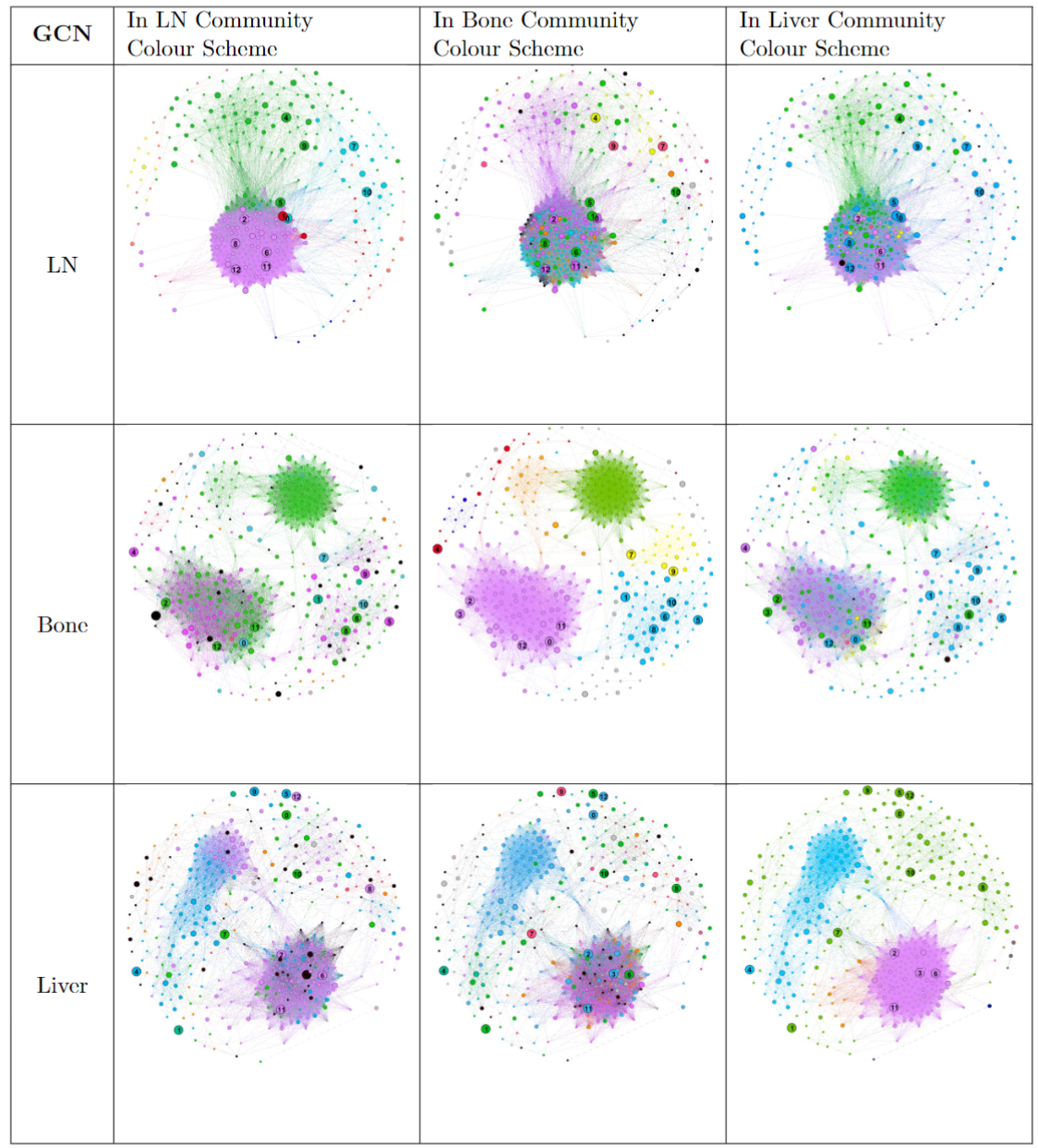}
\caption{\textit{LN, Bone and Liver GCNs visualization in different GCN community colour schemes. }}
\label{tab:colcomp}
\end{figure}

We made a summarization of the All-patient GCN, LN GCN, Bone GCN and Liver GCN and community details in table \ref{tab:comsum}.  To be noticed that, since we are interested in the distribution of the SET\_13 genes among the communities, we assign index numbers from 0 to 12 for each of the 13 key genes. In table \ref{tab:comsum}, we also provided a list of the key genes' indexes in each community in round bracket.\\

To compare the gene communities in different prostate cancer secondary site GCNs, we applied LN, Bone and Liver GCN community colour schemes for each of the GCNs. For the graphs in table \ref{tab:colcomp}, each row is community visualisations for one prostate cancer secondary site GCN in different GCNs’ community colour schemes. The layout of GCNs in same row keep the same, but genes may belong to different communities. However, each column is community visualisations for each GCN in one GCN’s community colour schemes. The graphs in the diagonal represent the GCNs in itself community colour scheme. We also use different node size to denote the importance of genes. The most important genes are SET\_13 genes, which are in the largest node size, and also annotated the index number of the 13 genes on each node. The second largest nodes are genes in SET\_34 but not in SET\_13, and the third largest nodes are genes in SET\_133 but not in SET\_34. The smallest nodes are genes in SET\_559 but not in SET\_133. A larger view of these GCN visualisations are also added in Appendix.\\

\section{Evaluation}
We can find there are strong gene community structures in these GCNs, particularly in the bone and liver GCNs. By altering community colour schemes with each GCN, we can also see that gene communities that occur in one GCN often occur in another GCN, which is strong evidence that gene community members tend to co-express.\\

However, we also noted that some genes went into different gene communities in different GCNs and there was some variation in the members that made up each community. This may be because the same genes play different roles in different kinds of metastatic cancer cells. In other words, different types of prostate cancer cells have different co-expression community structures, which allow the cells to synthesise different proteins and thus determine differences in cellular properties.\\

By deeply analysing the distribution of SET\_13 genes in the communities of different GCNs, we found that the key genes tended to be in different gene communities in different GCNs. For example, SET\_13 key genes with indexes 4, 5 and 9 were in the same community in the LN GCN but in three different communities in the Bone GCN. While in the Liver GCN, key genes 5 and 9 were in the same community and 4 was in another community. This apparent difference in the distribution of key genes could explain the effectiveness of the machine learning model.\\

\chapter{Discussion}
\section{Achievements}
In this interdisciplinary study, we have used data science, machine learning and network science methods to perform genetic analysis of metastatic prostate cancer and achieved promising results. In Chapter 5, we perform a thorough analysis of the raw dataset and then propose a general approach to pre-process the gene expression dataset. We filtered out 2 important gene sets, SET\_559 and SET\_133, using only the data analysis method. In Chapter 6, we further extracted two key gene sets, SET\_34 and SET\_13, from SET\_133 using the XGBoost model and trained a high-performance classification model using these genes. In Chapter 7, with intuition from biology and network science, we constructed gene co-expression networks for LN, Bone and Liver tumour samples respectively, and then performed network analysis and community detection on these networks. The results show that prostate cancer related genes have a strong gene community structure in a variety of metastatic cancer cells, and that these gene communities may play different roles, leading to cancer metastasis to different secondary sites.\\

\section{Methodological Contribution }
\subsection{Originality}
\quad\\
\textbf{(i) XGBoost model} 

In much of the literature on cancer cell classification and prediction based on sample gene expression values, machine learning models such as neural networks or support vector machine models are used. While these powerful models can produce accurate results by discovering underlying patterns behind the data, it suffers from interpretability problems. In this work, we used a simple decision tree-based model called XGBoost, as it supports scoring the importance of features. With the help of the feature importance scoring function of the XGBoost model, we successfully extracted several very important genes and achieved a high classification performance.\\

\quad
\textbf{(ii) Network analysis} 

Another unique aspect of this work is that we constructed multiple GCNs, each for one type of prostate cancer secondary site. Most literature analyses only one GCN, and a single network may provide only limited or unclear information. In this work, however, we compare and contrast the LN, Bone and Liver GCNs and attempt to discover new knowledge and patterns from them. We compare these GCNs by their network characteristics, network layout and community structure, and use visualisation to highlight these differences in a more engaging way. We also delved into the differences in gene communities across GCNs by tracking key genes.\\

\quad\\
\textbf{(ii) Ensemble methodology} \\

While much of the literature has been devoted to the genetic analysis of cancer using pure analysis methods, machine learning methods and gene co-expression network analysis methods respectively, very little work has combined them together to solve a complex problem. The most prominent approach in our work is to combine these three methods by taking advantage of their strengths.\\

In the preliminary work, we did a solid data analysis of the original dataset, and this process gave us a concrete understanding of the dataset, which will in a way make our next work easier and more sensible. We then used data analysis methods to pre-filter a number of genes to reduce the number of features used in the machine learning model, thereby improving the performance of the machine learning methods. Then, although with the help of powerful machine learning methods, we further extracted key genes and were able to accurately classify different secondary cells, we still did not know how genes were co-expressed at different metastatic sites. Therefore, we constructed GCNs for network analysis to reveal the co-expression patterns of genes. Also, with the help of machine learning results, we could track the different roles of key genes in different GCNs.\\

Through this research experience, we have gained a deepper understanding that each method has its advantages and disadvantages. It is very hard to solve a complex problem using only one method, but we can use the advantages of multiple methods to discover new insights to solve the problem. These methods are like rings in a chain, and the trickiest part is not how to use individual methods, but how to use the strengths of one method to complement the weaknesses of another, and find a way to bring these methods together to solve a complex problem.\\

\subsection{Transferability}
The methods used in this project are transferable, which means that the genetic analysis methods covered in this paper can also be applied to other gene expression datasets, and even other cancer research tasks. Although the current study was aimed at metastatic prostate cancer, none of our analytical methods were designed specifically for metastatic prostate cancer. Only the basic biology associated with cancer was considered in the methodological design.

\section{Limitation}
The pre-processed dataset used in this project contains only 231 samples from 3 types of patients. Also the sample size was very unevenly distributed between these 3 types. The small sample size and the few types of metastatic prostate cancer secondary sites may make the results of this study less generalisable. We therefore found another dataset of gene expression values for metastatic prostate cancer. However, due to the limited time available for this study, we did not validate the performance and compatibility of our trained XGBoost model, and the consistency of our GCNs on this dataset.\\

In the GCN analysis section, we only used the non-overlapping community detection algorithm to discover co-expressed genes. However, in terms of biology, some genes may be involved in the synthesis of multiple proteins, which means that there may be overlapping communities in the GCN. Therefore, it makes more sense to use an overlapping community detection algorithm instead. However, there are many overlapping community detection algorithms available, each may be suitable for a different application scenario. It would take a lot of time to investigate which algorithm is appropriate, and this may be beyond the scope of this project. There is also the fact that the analysis of overlapping communities is much more complex. As this work is a preliminary study of the genetic analysis of prostate cancer and time is limited, we have made a trade-off here by focusing only on detection of non-overlapping communities to simplify the problem.\\

\chapter{Future Work}
\section{ML Models}
Many publications mention that SVM models may be a good choice for tasks such as cancer classification. SVM with recursive feature selection can also be effective for selecting key genes. Therefore, in the next work, other feasible machine learning model solutions including SVM can be implemented and it would be interesting to compare these models together with XGBoost models and conclude which one works best. And, with a new dataset of gene expression values for metastatic prostate cancer recently collected by the Institute of Cancer Research in London, we can next train these machine models on the existing dataset used in this study, and then validate all of our trained models on this new dataset in order to verify which one is the best in terms of compatibility and performance.\\

\section{Gene Co-expression Network}
As we can see in the GCN visualisations, many communities have a large size, which could contain too many genes may not necessarily have co-expression relationship. So, in next step, we could do sub-community detection on large communities to further break the GCN into relatively small pieces. And then compare the gene communities in different GCNs, we may spot communities in different GCNs more consistent. \\

In addition, it is of considerable significance to do overlapping community detection in future work, as overlapping community detection is more closely related to the significance of gene co-expression in biology. Furthermore, this contribution will be of great help in future studies to develop drugs related to prostate cancer metastasis. For genes that are in overlapping regions of multiple communities, the drugs made may need to avoid affecting these genes as much as possible. This is because the presence of these genes in multiple co-expressed gene communities implies that if a drug has an effect on such genes, it may affect the cell's synthesis of multiple functional proteins, producing some complex side effects.

\appendix


\chapter{Model hyperparameters}

\begin{table}[h!]
\begin{tabular}{|ll|ll|}
\hline
\textbf{Hyperparameters} & \textbf{Value} & \textbf{Hyperparameters} & \textbf{Value} \\ \hline
base\_score & 0.5 & booster & gbtree \\ \hline
colsample\_bylevel{\color[HTML]{FFFFFF} .}{\color[HTML]{FFFFFF} .} & 1 & colsample\_bynode & 1 \\ \hline
colsample\_bytree & 1 & seed & 0 \\ \hline
learning\_rate & 0.1 & gamma & 0 \\ \hline
max\_depth & 3 & max\_delta\_step & 0 \\ \hline
missing & None & min\_child\_weight & 1 \\ \hline
nthread & 1 & n\_estimators & 100 \\ \hline
reg\_alpha & 0 & objective & binary:logistic \\ \hline
scale\_pos\_weight & 1 & reg\_lambda & 1 \\ \hline
subsample & 1 & verbosity & 1 \\ \hline
\end{tabular}
\caption{XGBoost model Hyperparameters setting}
\label{tab:my-table}
\end{table}

{\color[HTML]{FFFFFF} .}
\\
{\color[HTML]{FFFFFF} .}
\\
{\color[HTML]{FFFFFF} .}
\\
{\color[HTML]{FFFFFF} .}
\\
{\color[HTML]{FFFFFF} .}
\\
{\color[HTML]{FFFFFF} .}
\\
{\color[HTML]{FFFFFF} .}
\\
{\color[HTML]{FFFFFF} .}

\chapter{Figures}
\section{Effective of mask selection method}

\begin{figure}[h!]
\centering
\includegraphics[scale=0.7]{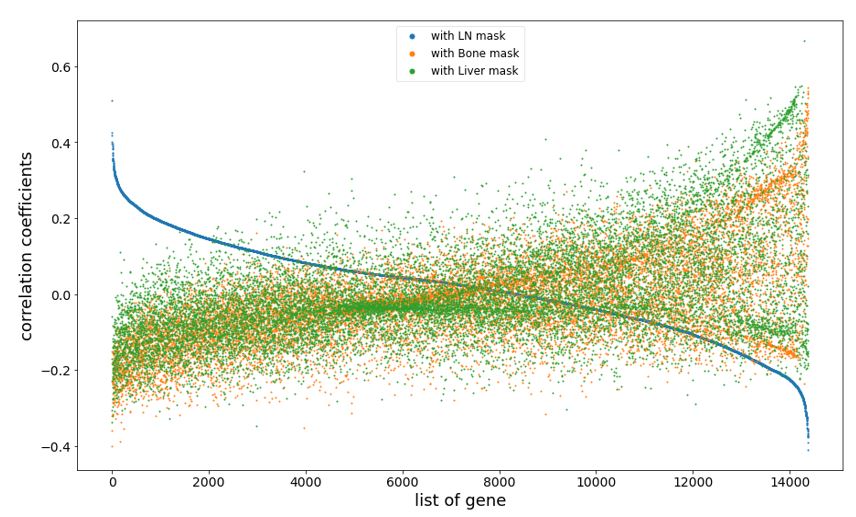}
\caption{X are 14,379 genes and Y are 3 correlation values with each of the LN,Bone and Liver metastases masks in different colours.  Sort the genes in descending order according to the correlation value with the LN metastases mask.}
\end{figure}

\begin{figure}[h!]
\centering
\includegraphics[scale=0.7]{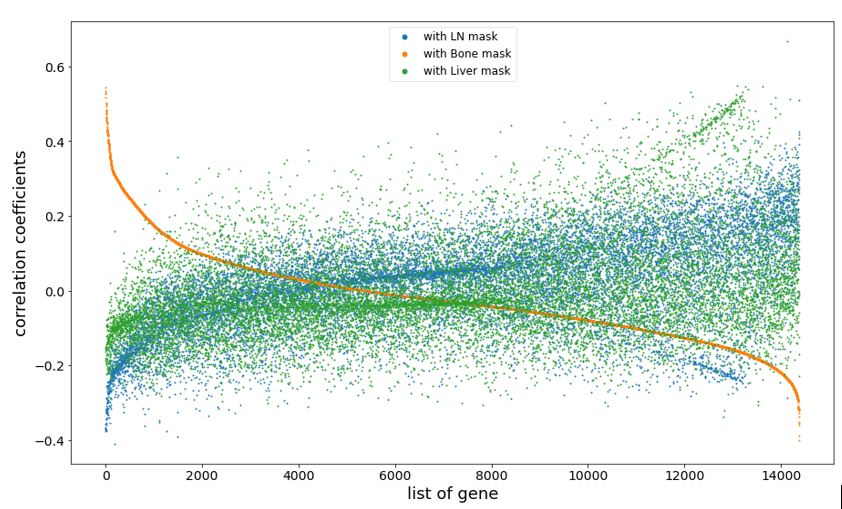}
\caption{X are 14,379 genes and Y are 3 correlation values with each of the LN,Bone and Liver metastases masks in different colours.  Sort the genes in descending order according to the correlation value with the LN metastases mask.}
\end{figure}

\begin{figure}[h!]
\centering
\includegraphics[scale=0.7]{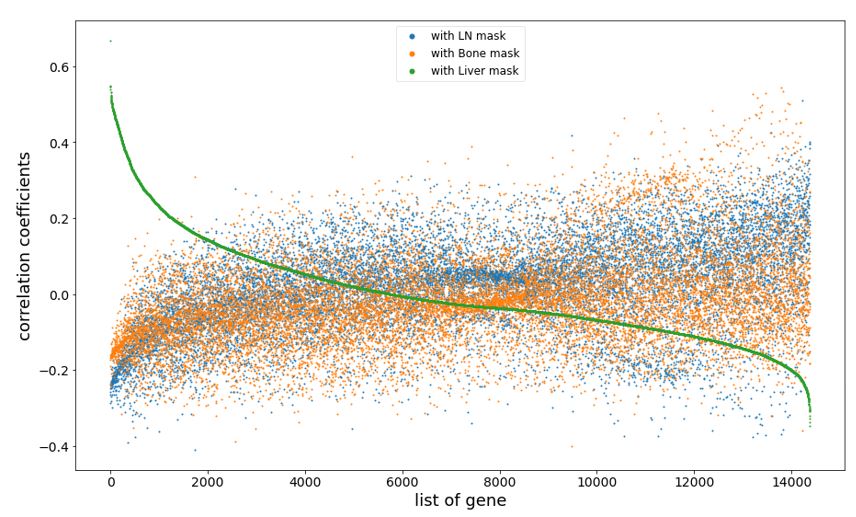}
\caption{X are 14,379 genes and Y are 3 correlation values with each of the LN,Bone and Liver metastases masks in different colours.  Sort the genes in descending order according to the correlation value with the Liver metastases mask.}
\end{figure}

\newpage

\section{Gene Co-expression Networks Visualisation}

\begin{figure}[h!]
\centering
\includegraphics[scale=0.5]{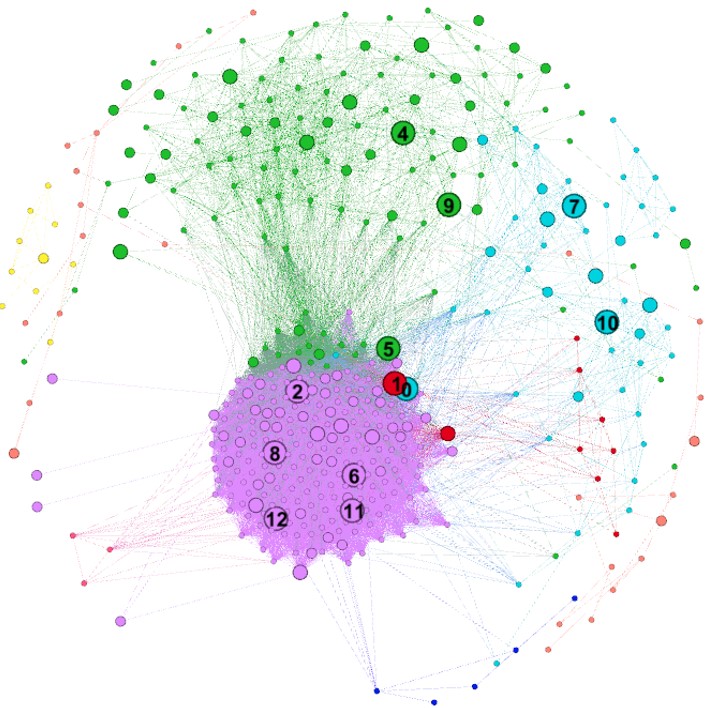}
\caption{LN GCN }
\end{figure}

\begin{figure}[h!]
\centering
\includegraphics[scale=0.5]{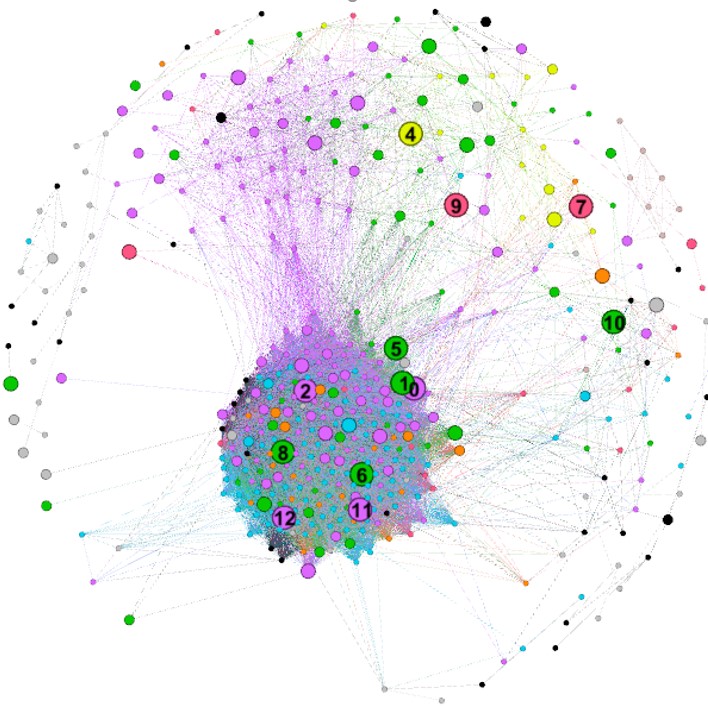}
\caption{LN GCN coloured by Bone GCN's community colour scheme}
\end{figure}

\begin{figure}[h!]
\centering
\includegraphics[scale=0.5]{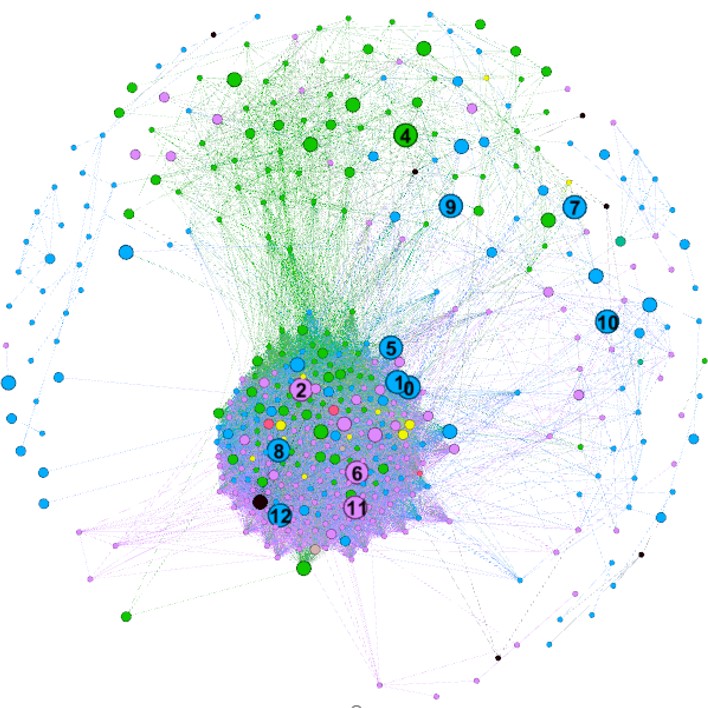}
\caption{LN GCN coloured by Liver GCN's community colour scheme}
\end{figure}

\begin{figure}[h!]
\centering
\includegraphics[scale=0.5]{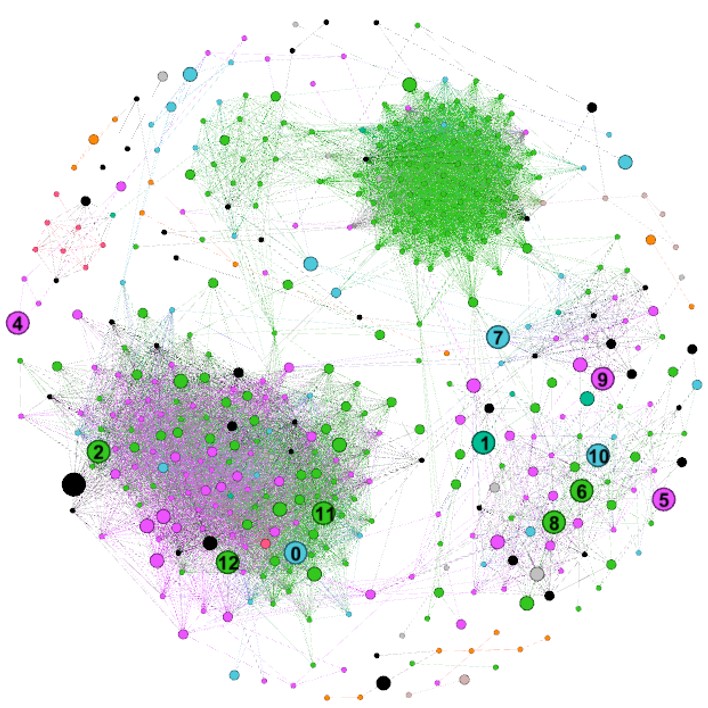}
\caption{Bone GCN coloured by Ln GCN's community colour scheme}
\end{figure}

\begin{figure}[h!]
\centering
\includegraphics[scale=0.5]{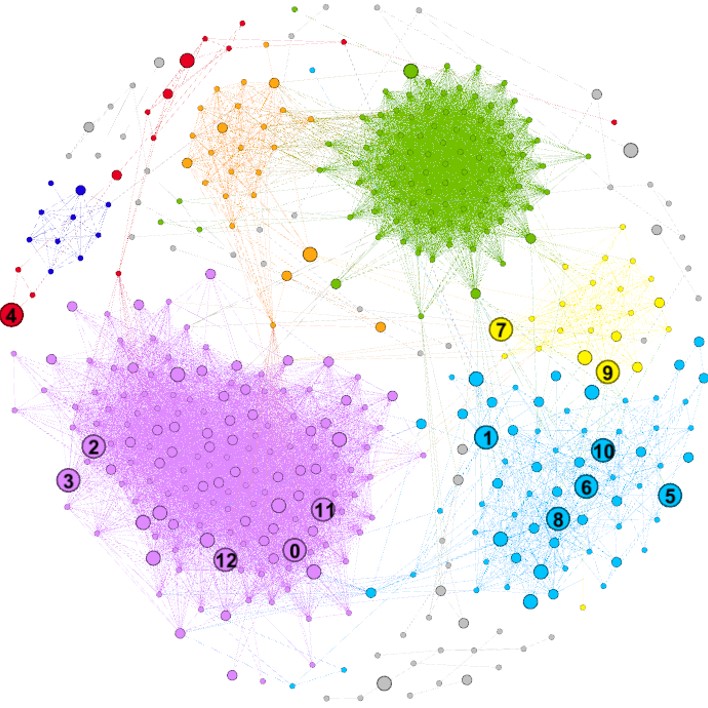}
\caption{Bone GCN}
\end{figure}

\begin{figure}[h!]
\centering
\includegraphics[scale=0.5]{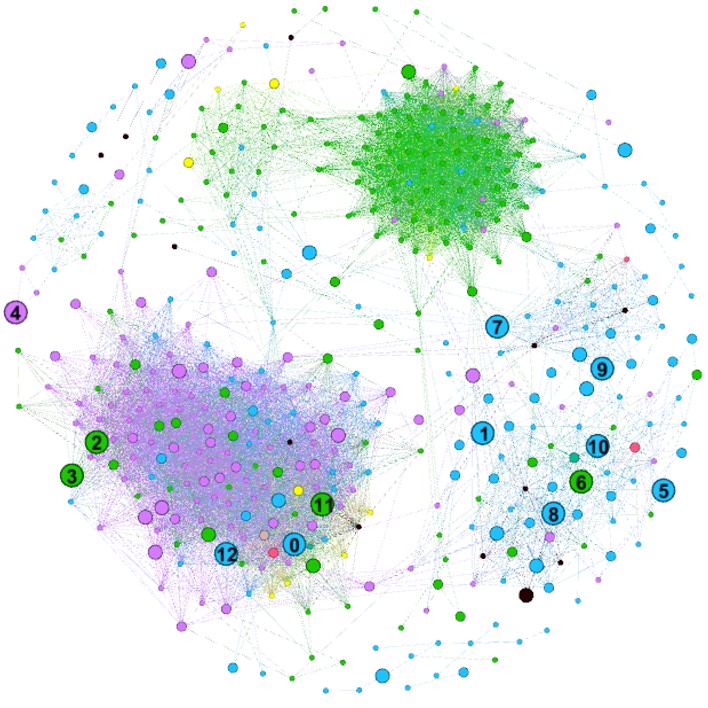}
\caption{Bone GCN coloured by Liver GCN's community colour scheme}
\end{figure}

\begin{figure}[h!]
\centering
\includegraphics[scale=0.5]{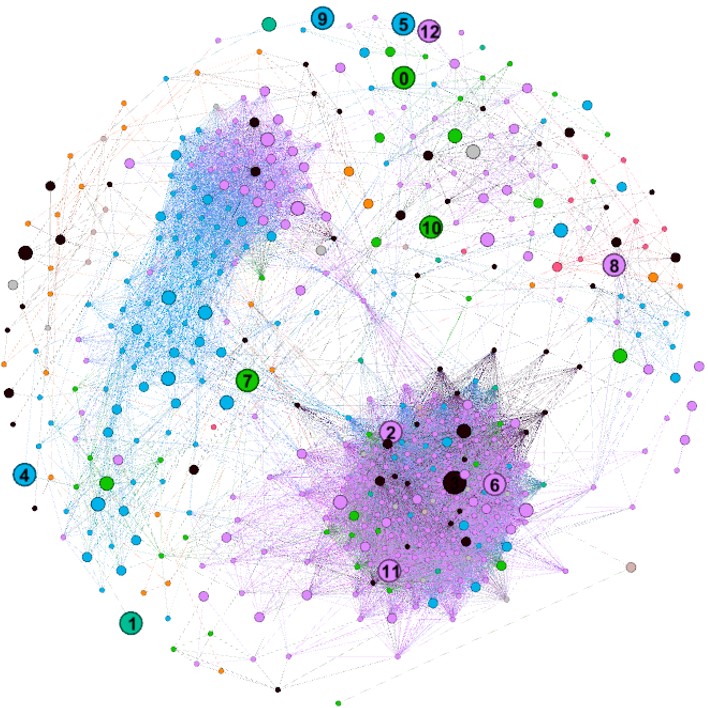}
\caption{Liver GCN coloured by LN GCN's community colour scheme}
\end{figure}

\begin{figure}[h!]
\centering
\includegraphics[scale=0.5]{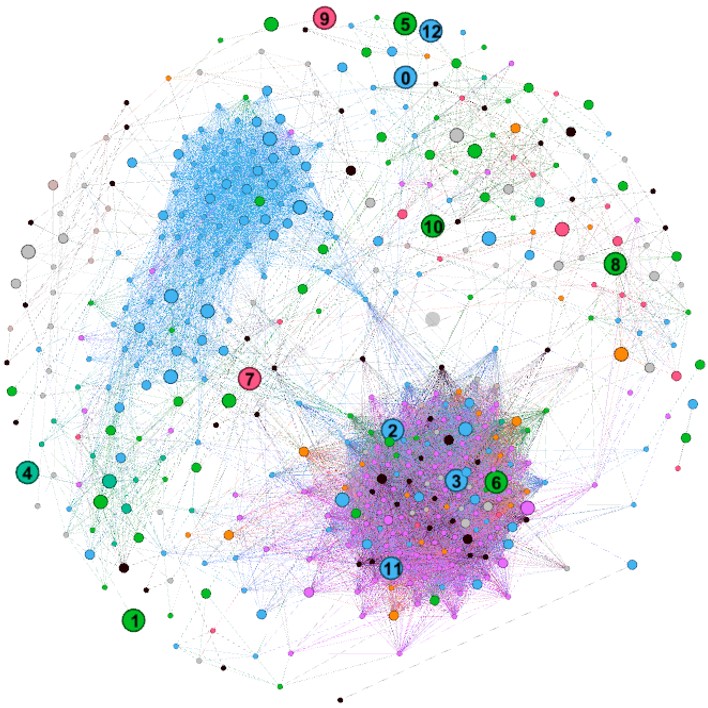}
\caption{Liver GCN coloured by Bone GCN's community colour scheme}
\end{figure}

\begin{figure}[h!]
\centering
\includegraphics[scale=0.5]{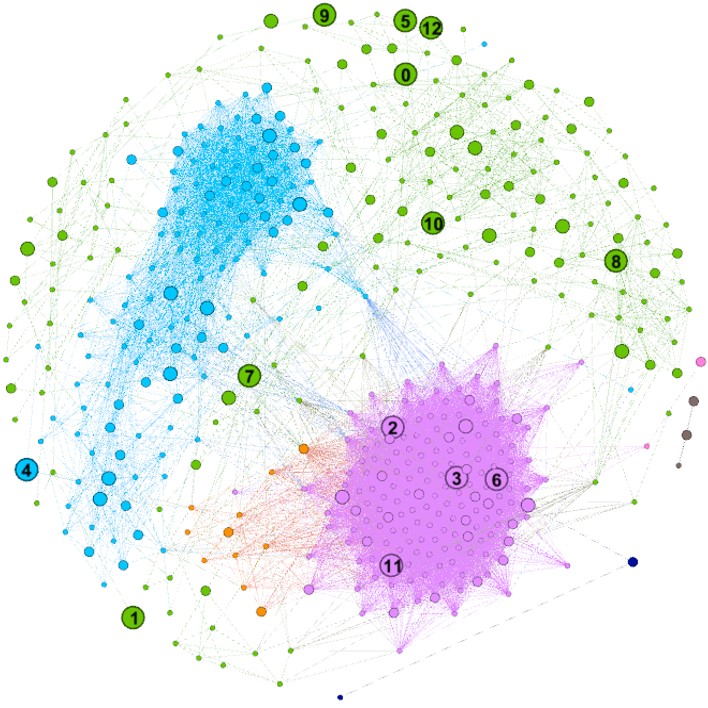}
\caption{Liver GCN}
\end{figure}

\chapter{Code and Dataset}
The code and datasets related to this study are all available at: \\ \textbf{\href{https://github.com/zcablii/cancer_project}{https://github.com/zcablii/cancer\_project}}.

\end{document}